\documentclass[a4paper,11pt]{article}
\pdfoutput=1 

\usepackage{jcappub} 
\usepackage{graphicx,color}
\usepackage{amsmath}
\usepackage{booktabs}
\usepackage{hyperref}
\usepackage[normalem]{ulem}
\usepackage[T1]{fontenc} 

\newcommand{\fermi}{\emph{Fermi}-LAT }
\newcommand{\dw}{\text{d} w}
\newcommand{\ie}{\emph{i.e.}\ }
\newcommand{\eg}{\emph{e.g.}\ }

\title{Classification of \emph{Fermi}-LAT blazars with Bayesian neural networks}
\author[a]{Anja Butter,}
\author[b]{Thorben Finke,}
\author[b]{Felicitas Keil,}
\author[b]{Michael Kr\"amer,}
\author[b]{Silvia Manconi}

\affiliation[a]{Institut f\"ur Theoretische Physik, Universit\"at Heidelberg, Germany}
\affiliation[b]{Institute for Theoretical Particle Physics and Cosmology, RWTH Aachen University, D-52056 Aachen, Germany}

\emailAdd{manconi@physik.rwth-aachen.de}
\arxivnumber{TTK-21-51}

\abstract{
The use of Bayesian neural networks is a novel approach for the classification of $\gamma$-ray sources. We focus on the classification of \fermi blazar candidates, which can be divided into BL Lacertae objects and Flat Spectrum Radio Quasars. 
In contrast to conventional dense networks, Bayesian neural networks provide a reliable estimate of the uncertainty of the network predictions. We explore the correspondence between conventional and Bayesian neural networks and the effect of data augmentation. We find that Bayesian neural networks provide a robust classifier with reliable uncertainty estimates and are particularly well suited for classification problems that are based on comparatively small and imbalanced data sets. The results of our blazar candidate classification are valuable input for population studies aimed at constraining the blazar luminosity function and to guide future observational campaigns.}

\begin{document}
\maketitle
\flushbottom

\section{Introduction}\label{sec:intro}
Since the first detailed observations of the $\gamma$-ray sky at GeV energies made by EGRET \citep{1999ApJS..123...79H}, source catalogs have collected a variety of $\gamma$-ray emitting objects in our Galaxy and beyond. After more than ten years of data taking, the \emph{Fermi}-Large Area Telescope (\emph{Fermi}-LAT), has detected almost six thousand point-like $\gamma$-ray sources~\cite{Ballet:2020hze}. The majority of the observed sources are blazars, \ie Active Galactic Nuclei (AGN) with a jet of outflows pointing towards the line of sight. Blazars can be divided further into BL Lacertae objects (BLL) and Flat Spectrum Radio Quasars (FSRQ) based on their spectral properties~\cite{Fermi-LAT:2019pir}. 

About one third of the blazar candidates detected by \fermi\ so far are of uncertain type. A reliable classification of blazar candidates would be important for population studies, which aim to improve our understanding of the blazar sequence~\cite{Ghisellini:2017ico}, and which allow to quantify the contribution of various sources to the extragalactic $\gamma$-ray background. Population studies may thus help to constrain possible exotic $\gamma$-ray signatures from \eg dark matter annihilations \cite{2015PhR...598....1F}. Furthermore, they are crucial for the  interpretation of astrophysical neutrinos as observed by IceCube \cite{Abbasi:2021jyg}, which could partially originate from blazars \cite{Giommi:2020hbx}.

In order to obtain an unambiguous blazar classification, extensive optical spectroscopy and multiwavelength observations are required, see \eg refs.~\cite{deMenezes:2020yru,Rajagopal:2021dvq}, which are however time-consuming and costly. Moreover, the increasing number of blazar candidates detected by \fermi poses a challenge for a timely follow-up observation of each source. Accordingly, the number of blazars of uncertain type has increased from about $13\%$ in the first \fermi catalog~\cite{Abdo_2010} to about $40\%$ in the most recent edition~\cite{Ballet:2020hze}. Thus, an efficient method is required that provides a first classification of blazar candidate sources to tailor further observational campaigns.

Machine learning methods, including neural networks, have been used for the classification of $\gamma$-ray sources in various analyses, ranging from the identification of AGN and pulsar candidates~\cite{Mirabal_2016,Saz_Parkinson_2016,Luo_2020,Finke:2020nrx,Bhat:2021wtb,Panes:2021zig} and of blazars~\cite{Doert_2014,Chiaro:2016noj,Salvetti_2017,Kovacevic:2020sly,Kerby:2021hep} to the search for new exotic source classes such as dark matter subhalos \cite{Mirabal:2012em}. 
However, apart from the usual performance tests done on the training and testing data sets, it is not clear in general how to estimate the uncertainty associated with the machine learning output. Bayesian neural networks (BNNs) \cite{MacKay1992, neal2012bayesian, Gal2016Uncertainty} provide a natural formalism to quantify uncertainties associated to neural network predictions. They have been employed in astroparticle physics and high-energy physics, for example for jet classification~\cite{Bollweg:2019skg}, jet energy calibration~\cite{Kasieczka:2020vlh}, event generation~\cite{Bellagente:2021yyh, Butter:2021csz}, supernovae classification~\cite{2020MNRAS.491.4277M}, or to investigate the nature of the Galactic center excess~\cite{List:2020mzd}, but thus far not for the classification of $\gamma$-ray sources.

We use neural networks trained on the energy spectra of known BLL and FSRQ to classify  blazars of uncertain type. Using the energy dependent flux instead of derived features as input for neural network classifiers has already been demonstrated to be a powerful method for various $\gamma$-ray source classification tasks \cite{Kovacevic:2020sly,Finke:2020nrx}. The novel aspect of our work is the use of BNNs, which allow us to quantify the uncertainty of the classification prediction. Such uncertainty estimates provide insight into the performance and reliability of different machine learning algorithms, and they may help to select the most relevant blazar candidates for subsequent observations. As the data set available for neural network training includes only about 2000 blazars, we also explore the effect of data augmentation techniques, which are often  used to improve the performance of neural networks when only small data sets are available.

The paper is organized as follows. The data set is introduced in section~\ref{sec:data}. Our classification methods, including the main neural network architectures, the training and testing strategies as well as the exploration of data augmentation are presented in section~\ref{sec:methods}. In section~\ref{sec:results} we discuss the classification performance, a cross-matching procedure using an older \fermi\ data set, and we provide our predictions for the blazars of uncertain type  in the current \fermi\ catalog. We conclude in section~\ref{sec:conclusions}. Further results are presented in the appendix. A link to a repository containing ancillary files with our classification results is provided at the end of the manuscript. 

\section{Blazars in the Fermi-LAT catalog} \label{sec:data}

\begin{figure}
	\center
	\includegraphics[width=.7\textwidth]{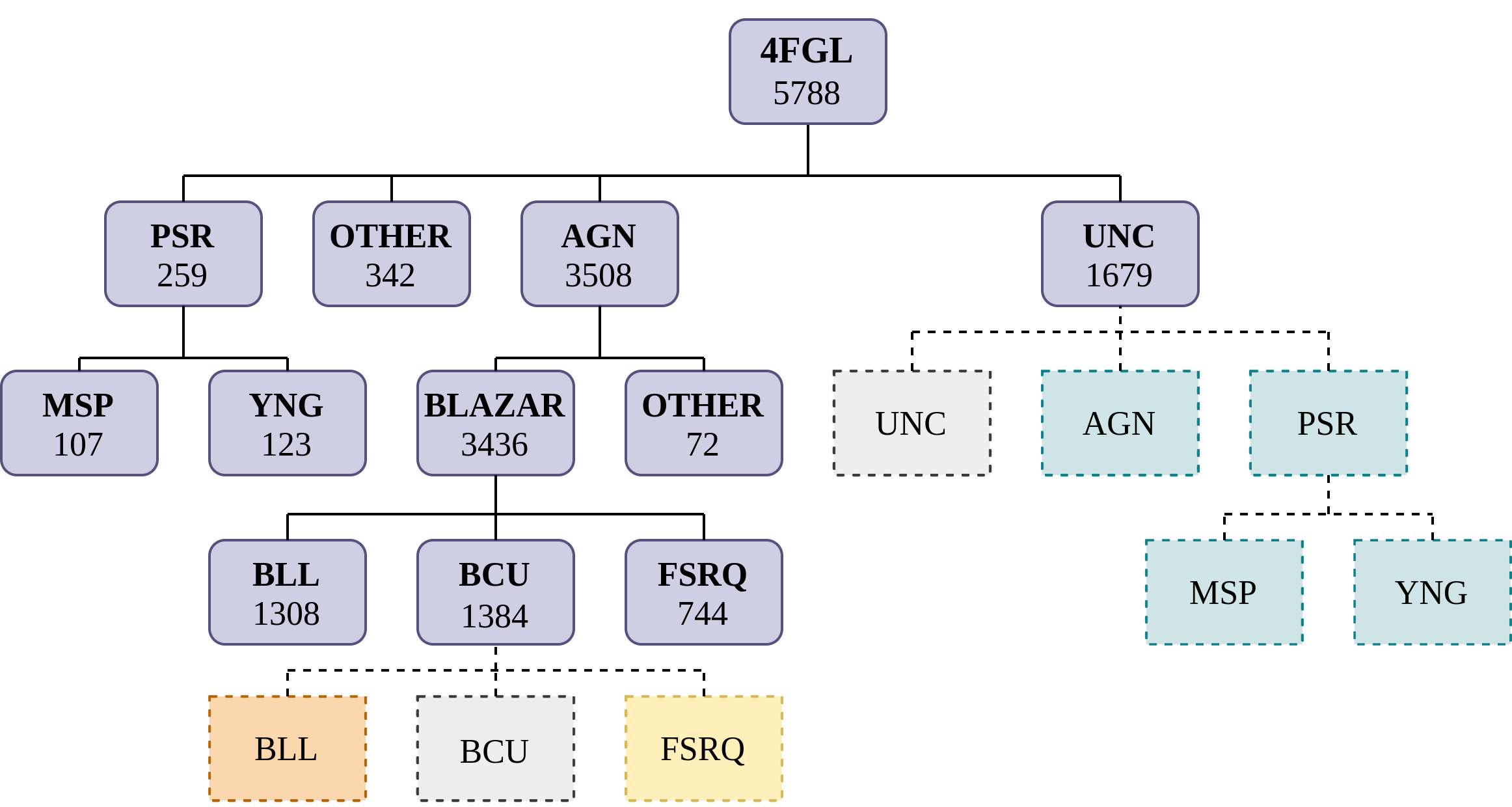}
	\caption{
	Classification tree for the 5788 sources in the \fermi 4FGL-DR2 catalog.
	The violet boxes (connected by solid lines) denote the classification provided in the catalog,
	while the dashed boxes indicate the predictions which can be obtained using \eg machine learning classifiers.
	The cyan boxes (connected by dashed lines) indicate the predictions for the classes of UNC sources as performed in \cite{Finke:2020nrx}.
	The focus of this work is the classification of BCU into BLL and FSRQ (orange and yellow boxes, see text for more details).}
	\label{fig:classtree}
\end{figure}

Sources observed by \fermi are characterized according to their position and spectral characteristics. 
They are classified as \emph{identified} 
if there is a correlated timing signature at different wavelengths, or \emph{associated} 
if only a positional coincidence with a counterpart source at other wavelengths has been found. 
The fourth \fermi $\gamma$-ray catalog, Data Release~2, (4FGL-DR2) is the most recent and comprehensive catalog of $\gamma$-ray emitting sources. It is based on ten years of data with $\gamma$-ray energies from 50~MeV to 1~TeV \citep{Ballet:2020hze} and uses the same analysis methods as the previous 4FGL catalog \citep{Fermi-LAT:2019yla}.
The classification tree for the 4FGL-DR2 \fermi catalog is shown in figure~\ref{fig:classtree} and discussed in more detail below.

The main source classes in the current catalog are Active Galactic Nuclei (AGN) and pulsars (PSR).
PSR are rapidly rotating, highly magnetized neutron stars surrounded by a plasma magnetosphere.
AGN is a generic name for sources correlated with jets originating from a supermassive black hole at the center of a galaxy, emitting electromagnetic radiation in a broad range, from radio frequencies to $\gamma$-rays with TeV energies. More than 90\% of the AGN detected by \fermi  are blazars, \ie AGN where the jet is pointing towards the line of sight.
Other classes of sources include more rare types of galaxies or Galactic emitters such as supernova remnants.
A total of 1679 sources (about one third) in the 4FGL-DR2 catalog remains unclassified (UNC).

In this work we will focus on the subclassification of blazars, which are divided into BL Lacs (BLL), Flat Spectrum Radio Quasars (FSRQ), and blazars of uncertain type (BCU), see also the dedicated AGN catalog \cite{Fermi-LAT:2019pir} and section~\ref{subsec:bla}. Our classification task is illustrated by the 
final branch on the left of figure~\ref{fig:classtree}.
As demonstrated in previous works, the separation of blazars into BLL and FRSQ is a task which can be addressed very well with machine learning techniques. 
In particular, the classification of $\gamma$-ray blazar candidates can be performed based on a reasonable statistics, as blazars are more than half of the total labeled sources in \fermi catalogs. Moreover, compared to the AGN vs.\ PSR classification task, a rather balanced number of objects in each class (1308 vs.\ 744) is available for training and testing.
Note that the branch on the right of the classification tree, figure~\ref{fig:classtree}, is the classification task discussed for example in \cite{Finke:2020nrx}, taking the UNC sources and searching for AGN or PSR candidates.

\subsection{Characteristics of FSRQ and BLL blazars}\label{subsec:bla}

Our goal is to identify BLL and FSRQ candidates among the 1384 4FGL-DR2 blazars classified as uncertain type (BCU).
According to the unified classification scheme of AGN \citep{Padovani:2017zpf}, they all correspond to the nucleus of external galaxies with jet and tori emitting both thermal and non-thermal emissions.
Different types of AGN (such as the BLL and FSRQ) are then separated according to their observational properties. 
The classification of BLL and FSRQ is based on the observational characteristics of their emission lines and continuum emission, and might not always correspond to a distinctive physical difference between the two classes~\citep{Ghisellini:2017ico}.
In general, observations suggest that FSRQ have strong external radiation fields, strong and broad optical emission lines as well as dust tori, while BLL have weaker emission lines and sometimes absorption features.
Thus the observed populations may not always represent the real cosmic abundances, but may rather be driven by observational bias, see \eg the discussion in ref.~\cite{Padovani:2017zpf}.

In the $\gamma$-ray energy range, the mean spectral power law index $\Gamma_\gamma$ 
and the luminosity $L_\gamma$ are $\Gamma_\gamma \ge 2.2$ and $L_\gamma \ge 1039$~W for FSRQ and $\Gamma_\gamma \le 2.2$ and $L_\gamma \le 1039$~W for BLL.
The BCU are found to have
intermediate characteristics, notably in $\Gamma_\gamma$, indicating that they are likely a mixture of BLL and FSRQ.
Since these two classes exhibit different $\gamma$-ray emission characteristics, the classification of BCU as BLL or FSRQ is very important for population studies, see the detailed discussion in section~\ref{sec:pred}.

\subsection{Data Set for classification}
The \fermi 4FGL-DR2 catalog\footnote{\texttt{gll\_psc\_v27.fit}, publicly available at \url{https://fermi.gsfc.nasa.gov/ssc/data/access/lat/10yr_catalog/}} \citep{Ballet:2020hze} is the reference data set for our work.
We select both identified (reported as $\texttt{BLL, FSRQ}$ in the catalog column \texttt{CLASS1}) and associated BLL and FSRQ ($\texttt{bll, fsrq}$) from the labeled sources. 
This amounts to a total of 2052 sources (1308 BLL and 744 FSRQ) available to train and test our networks, see also figure~\ref{fig:classtree}.

The \fermi catalog provides 74 other features and measurements apart from the source positions and source class. 
We are interested in the energy spectrum, \ie the measured flux as a function of energy. 
As shown in ref.~\cite{Finke:2020nrx}, the $\gamma$-ray flux as a function of time (provided in yearly bins) adds little information for the classification.
The energy spectrum is fitted by the \fermi collaboration with different spectral forms, and best fit parameters are reported in the catalogs.
For example, the AGN energy spectrum is generally well reproduced by a power law with index between 1.5-3.
In this work, we want to use directly the energy spectrum (column labeled as \texttt{Flux\_Band} in the catalog) to extract as much information as possible from the actual measurements.
This approach has worked well for the classification of AGN vs.\ PSR \cite{Finke:2020nrx}, and should be suitable also for blazar classification, see \eg ref.~\cite{Kovacevic:2020sly}.
The 4FGL-DR2 catalog reports fluxes for seven energy bands, extending from 50 MeV to 300~GeV (1: 50--100 MeV; 2: 100--300 MeV; 3: 300 MeV--1 GeV; 4: 1--3 GeV; 5: 3--10 GeV; 6: 10--30 GeV; 7: 30--300 GeV \cite{Fermi-LAT:2019yla}).

\begin{figure}
\centering
\includegraphics[width=0.52\textwidth]{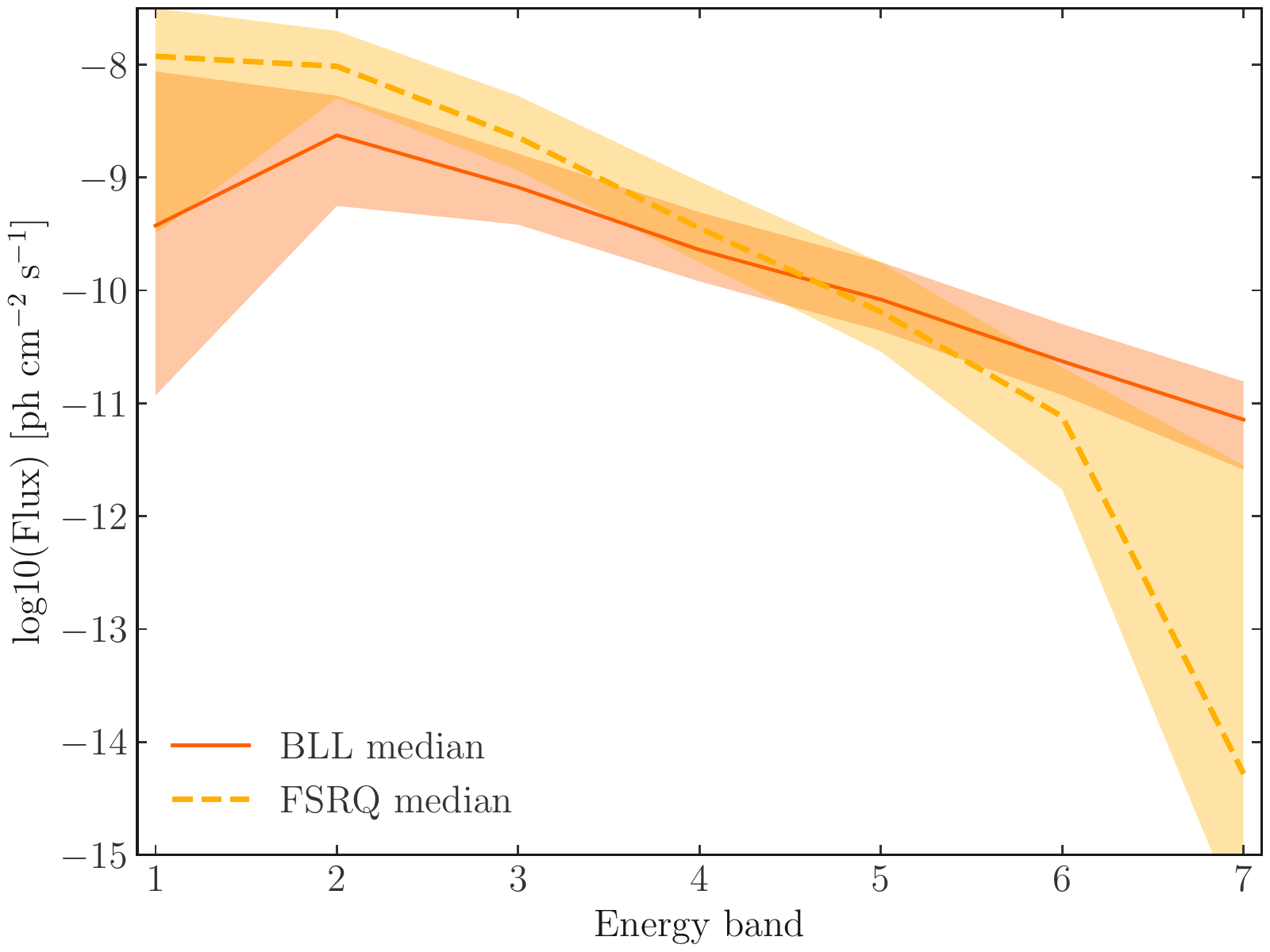}
\caption{Median energy spectrum of the BLL (solid line) and FSRQ (dashed line) blazar classes of the 4FGL-DR2 catalog used to train our networks (before preprocessing).
The lines correspond to mean spectra, while the bands represent the lower and upper quartile for each bin.
Energy bands extend from 50--100 MeV (band 1) to 30--300 GeV (band 7), see text for details.} \label{fig:spectra}
\end{figure}


The median energy spectra of the sources in the BLL (solid line) and FSRQ (dashed line) classes used to train our networks (before preprocessing, see section~\ref{sec:dnn}) are illustrated in figure~\ref{fig:spectra}.
The logarithm of the flux (in units of ph cm$^{-2}$ s$^{-1}$) is shown as a function of the energy bins specified above.
The median value is computed for each energy band separately, and the energy values for each bin are connected with lines for better visualization. 
The shaded bands show the lower and upper quartile with respect to the median of the two source classes for each energy bin.
We note that the BLL (FSRQ) are characterized by a larger variation towards the lower (upper) energy bin.
Most importantly, the difference in the mean spectral index ($\Gamma_{\gamma}\sim2$ for BLL, $\Gamma_{\gamma}\sim2.5$ for FSRQ) is clearly visible.

\section{Classification methods}\label{sec:methods}

Neural networks are powerful tools for the classification of $\gamma$-ray sources, see \eg refs.~\cite{Kovacevic:2020sly,Finke:2020nrx}. However, since the data set available for training and testing is comparatively small, standard neural networks are prone to overtraining and thus may generalize poorly. As we shall demonstrate, common data augmentation techniques cannot resolve this problem for our data set. Moreover, training a standard neural network repeatedly on the same training data in general leads to an overly optimistic estimate of the predictive uncertainty. We hence employ BNNs which are less susceptible to overtraining in general, and which are able to estimate the uncertainties on their prediction more reliably. In the following we introduce regular and Bayesian neural networks, describe the corresponding  architectures and our strategies for training and testing. We then analyze the performances of the regular and Bayesian neural networks using a toy data set, and explore the effect of data augmentation.
The networks used in this work are implemented using \textsc{Tensorflow 2.4.1}~\cite{tensorflow2015-whitepaper} and the built-in version of \textsc{Keras}~\cite{chollet2015keras}. For the BNN we exploit the implementation of probabilistic layers within \textsc{Tensorflow-Probability 0.12.1}~\cite{DBLP:journals/corr/abs-1711-10604}.

\subsection{Dense neural networks}\label{sec:dnn}
We first introduce a dense neural network (DNN) to establish a baseline for the classification of the $\gamma$-ray sources.
As demonstrated in ref.~\cite{Finke:2020nrx}, we can extract the information relevant for classification directly from the energy spectra, avoiding any potential loss of information in the construction of high-level observables.
The input of the network consists of seven energy bins with the corresponding logarithmic flux values as illustrated in figure~\ref{fig:spectra}.
To improve the convergence of the network training we perform data preprocessing by shifting the mean to zero and scaling the standard deviation to one for each flux bin individually.

The network architecture described below is very similar to the DNN presented in ref.~\cite{Finke:2020nrx} when using only the energy spectrum.
The input is processed by one dense layer with ReLU activation function followed by a softmax output layer with two nodes, corresponding to the respective class probabilities.
The training minimizes the binary cross entropy loss in combination with $L_2$-regularization and is optimized with ADAM \cite{adam}.
During the training of a DNN, individual weights and biases are trainable parameters within each hidden layer, and they are tuned to minimize the loss function.
Further details of the network architecture and training can be found in table~\ref{tab:param}.
Apart for the number of layers and the nodes of the dense network, we specify the number of training data the network uses in each step (batch size), the number of cycles through the full training data set (epochs), and the learning rate, which corresponds to the step size along the gradient while minimizing the loss function with gradient descent.
We performed a hyperparameter search exploring the variation of all parameters listed in table~\ref{tab:param}.
While the performance is robust against smaller changes in the batch size, epochs or learning rate, we observe that a DNN with too high expressivity is prone to overfitting the small data set.
We cannot tune the $L_2$-regularization term to prevent this overfitting as it is fixed according to the correspondence with the BNN (see section~\ref{sec:bnn_dnn_losses}). Thus, we keep the architecture shallow and use one hidden layer only.

The performance of the networks is evaluated using the standard 10-fold cross validation procedure, see ref.~\cite{Finke:2020nrx} for more details, and quantified by means of the Receiver operating characteristic (ROC) curve and the corresponding area under curve (AUC). We also provide the confusion matrix for some exemplary case.

The network architecture and hyperparameter settings, the data preprocessing, the training and testing strategies and the performance measures outlined above will be the same for the different networks discussed in the next sections.

\begin{table*}[b]
\caption{Hyperparameter settings for the DNN and BNN described in section~\ref{sec:methods}.}
\center
\begin{tabular}{lc}
\toprule
               &  DNN \&  BNN \\
\midrule
Layers         &    1   \\
Nodes          &    16  \\
Batch size     &    32  \\
Epochs         &    250 \\
Learning rate  &  $10^{-3}$\\ 
\bottomrule
\end{tabular}
 \label{tab:param}
\end{table*}

\subsection{Bayesian neural networks}
BNNs replace the individual weights of a DNN with weight distributions \cite{MacKay1992, neal2012bayesian, Gal2016Uncertainty}. 
For a fixed input, the evaluation of the weight distributions generates a distribution over probable output values.
The shape of the distribution allows us to assign an uncertainty to the prediction that captures the convergence as well as statistical uncertainties of the data set.
During training the BNN has to learn the true model posterior $p(w|D)$, \ie the distribution over network weights $w$ given a data set $D$.
Due to the complexity of the model, we cannot infer the true model posterior directly.
Instead, we approximate the posterior through variational weight distributions $q_\theta(w)$ with learnable parameters $\theta$.
We then minimize the difference between $q_\theta$ and $p$ through variation of the parameters $\theta$.
We use the KL-divergence as difference measure, which can be re-written as
\begin{align}
	{\rm KL}(q(w)||p(w|D))
	\notag &= \int \dw \;  q(w) \log \dfrac{q(w)}{p(w|D)}\\
	\notag &= \int \dw \;  q(w) \log \dfrac{q(w)}{p(D|w) p(w)} + \text{const}\\
	&= {\rm KL}(q(w) || p(w)) - \int \dw \; \ q(w) \log(p(D|w)) + \text{const} \text{ ,}
	\label{eq:KL1}
\end{align}
where we suppress the label $\theta$ for better readability.
The first term is data independent and quantifies the difference between the estimated posterior and the prior.
The second term estimates the log-likelihood of the data given the weight distribution. 
The third term originates from the prior over the data set itself, $p(D)$, which is independent of the weight parameters and hence irrelevant for the training.
If we take the average of eq.~\eqref{eq:KL1} over the training data set of size $N$, the KL-divergence reads
\begin{align}
	\frac{1}{N} {\rm KL}(q(w)||p(w|D))
	\notag &= \frac{1}{N} {\rm KL}(q(w) || p(w)) -  \frac{1}{N} \sum_{i=1}^{N} \log(p(x_i|w)))\\
	&\approx \frac{1}{N} {\rm KL}(q(w) || p(w)) - \frac{1}{M} \sum_{i=1}^{M} \log(p(x_i|w)) \text{ .}
	\label{eq:KL2}
\end{align}
The last step corresponds to the approximation of the sample average as an average over a batch of size $M$, corresponding to the loss calculated for a mini-batch update.
Furthermore, the final expression illustrates how larger data sets suppress the first term in eq.~\eqref{eq:KL1} and are hence able to mitigate the effect of the prior.

The approximate posterior ($q$) and the prior ($p$) are usually considered to be multivariate, diagonal normal distributions, with the posterior's parameterization being trainable.
As we use diagonal covariance matrices, we can parametrize both distributions with their respective mean values $\mu_{p/q}$ and their standard deviations $\sigma_{p/q}$.
The KL-divergence can then be calculated independently for each weight $w_i$ in a layer resulting in a regularization term of
\begin{equation}
	{\rm KL}(q(w) || p(w)) = \sum_{i} \log \frac{\sigma_{p,i}}{\sigma_{q,i}} + \frac{\sigma_{q,i}^2 + (\mu_{p,i} - \mu_{q,i})^2}{2 \sigma_{p,i}^2} - \frac{1}{2} \text{ ,}
	\label{eq:KL3}
\end{equation}
where the sum is taken over all weights. For practical purposes we can further simplify the contribution of the KL-divergence to the loss. Without loss of generality we can choose a normal distribution with vanishing expectation values $\mu_{p,i}=0$. Furthermore, we can neglect all constant terms like the logarithm of the prior widths. The simplified KL-divergence then reads
\begin{equation}
	{\rm KL}_{\text{simp}}(q(w) || p(w)) = 
	\sum_{i} 
	 \frac{\mu_{q,i}^2}{2 \sigma_{p,i}^2} 
	 + \frac{\sigma_{q,i}^2}{2 \sigma_{p,i}^2}
	 - \log \sigma_{q,i}\text{ ,}
	\label{eq:KL4}
\end{equation}
and is used in order to compute the regularization of the BNN.

\subsection{Correspondence between dense and Bayesian neural networks} \label{sec:bnn_dnn_losses}
Having introduced the BNN, we can estimate the uncertainty of the prediction for individual data points. An alternative approach to estimate the uncertainty of a prediction is to sample over the predictions of an ensemble of independently trained DNN.
We can compare the result of both approaches from a theoretical and applicational point of view. 

Since an individual DNN has no weight distribution we can identify its weights with the mean value of the BNN. As a result we can identify the first term in eq.~\eqref{eq:KL4} with a standard $L_2$-regularization, which sums up the network weights in quadrature,
\begin{equation}
	L_2
	= \lambda \sum_{i} w_{i}^2
	=\sum_{i} \frac{\mu_{q,i}^2}{2 \sigma_{p,i}^2} \text{ ,}
\end{equation}
once we set $\lambda = \frac{1}{2 \sigma_p^2}$.
Since the regularization term in eq.~\eqref{eq:KL4} has to be divided by the size of the training data set, the same prefactor needs to be applied to the $L_2$-regularization term in order to achieve comparable performance. The regularization helps to avoid overtraining and increases the variation in the prediction of different DNN for individual data points.

While this regularization term acts similarly in both settings there remain significant differences.
First of all, the weight distributions of an ensemble of DNN is not expected to follow a Gaussian, which is enforced in a BNN. 
Furthermore, the remaining terms in eq.~\eqref{eq:KL4} that take into account the width of the weight posterior lead to an additional regularization. These additional contributions to the loss originate naturally from the ansatz in eq.~\eqref{eq:KL1}, which take into account the limited size of the data set.

Before applying our networks to the 4FGL-DR2 data set, we want to explore this correspondence and its practical implications in a well-controlled environment using a toy model.
In order to stay close to the structure of the original data set we sample our toy data set from two multivariate, seven-dimensional Gaussian distributions.
The expectation values of the first distribution are set to decrease linearly from $1$ to $-1$, with corresponding standard deviations that decrease linearly from $1.1$ to $0.9$ along the seven bins/dimensions.
The order of the mean values and standard deviations of the second Gaussian are inverted, \ie the expectation values increase linearly from $-1$ to $1$ while the standard deviations increase from $0.9$ to $1.1$.
For our experiments we use the hyperparameters as given in table~\ref{tab:param}.
The size of the training data set is fixed to 1000 samples per distribution.
We train the BNN once and sample 100 times from the latent space of the weight distributions to obtain 100 predictions for each data point.
We want to compare the performance of the BNN to the predictions we obtain from an ensemble of 100 DNN trained with independent initialization.
In order to estimate the impact of reusing small training data sets, we distinguish two cases.
On the one hand all 100 DNN are trained on the same training data set, on the other hand we sample each of the 100 training data sets independently.
We compare the predictions of these three setups in terms of their mean prediction as well as their estimated uncertainties.

\begin{figure*}[h]
\centering
	\includegraphics[trim = 50 50 50 50, clip, width=1.05\textwidth]{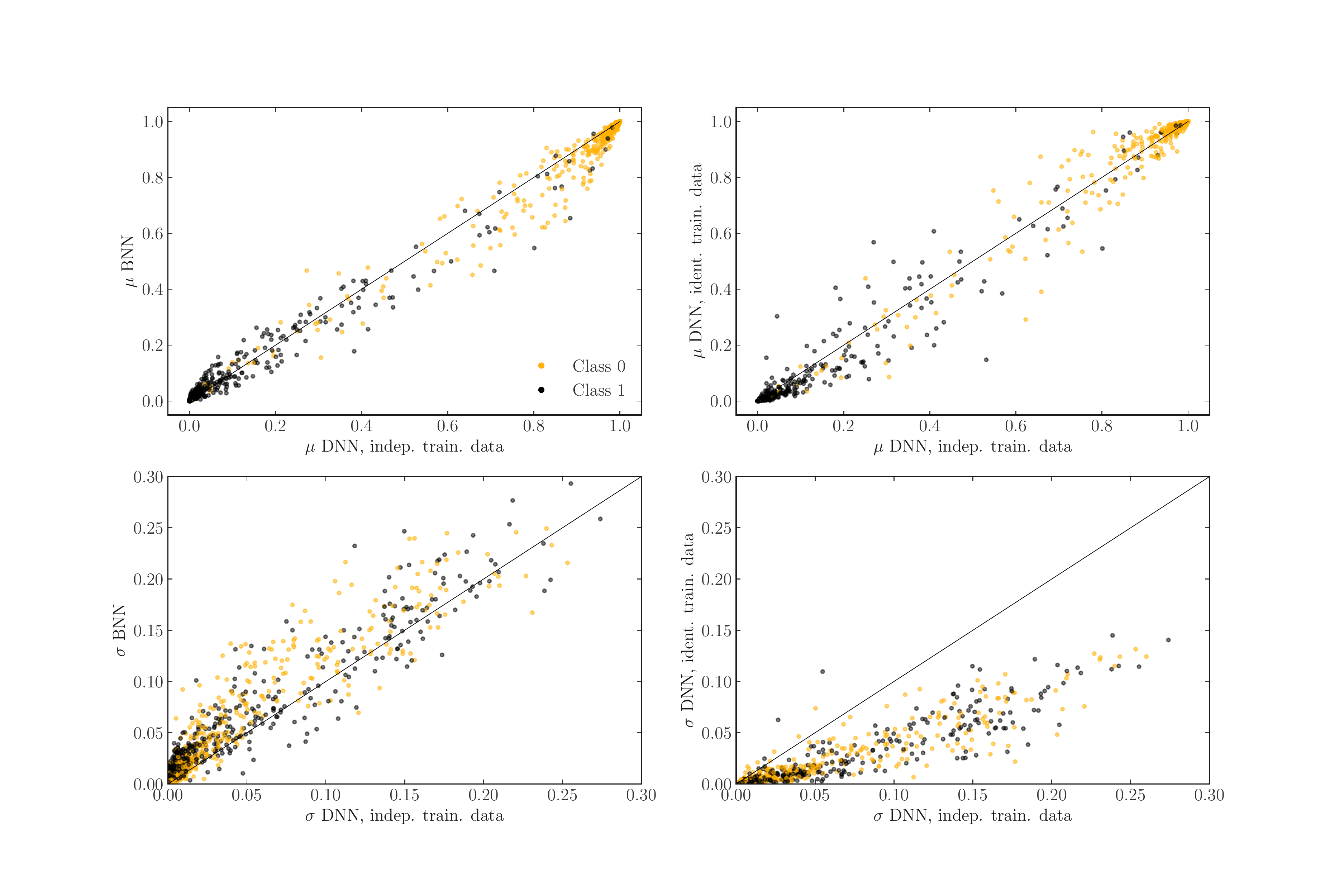}
	\caption{Correlation of the mean predictions ($\mu$, first row) and standard deviations ($\sigma$, second row) for different neural network architectures in our toy model setup (see section~\ref{sec:bnn_dnn_losses} for details). The results of the BNN (left column) and of 100 DNN trained on identical training sets (right column) are compared to the results of 100 DNN trained on independent training sets.}
	\label{fig:correlation_bnn_dnn}
\end{figure*}

In the upper row of figure~\ref{fig:correlation_bnn_dnn} we show the correlation of the mean predictions ($\mu$) for test samples from both classes between the BNN and the two DNN distributions.
In particular, we compare the 100 DNN trained on independent training data to the BNN (left panels) and to the DNN trained on identical training data (right panels). 
Both cases show a clear linear correlation between the mean predictions.
However, the correlation between BNN and the DNN trained on independent training data is much stronger as indicated by the small spread of results around the diagonal.
The lower panels of figure~\ref{fig:correlation_bnn_dnn} show the correlations of the standard deviation ($\sigma$) associated to the different predictions. 
Again we observe a linear correlation when comparing the standard deviation of the BNN with the DNN trained on independent samples.
On the other hand, training multiple DNN on the same data set significantly underestimates the uncertainties.
The variation of DNN trained on the same data set is only able to capture the stability of the convergence of the DNN to a local minimum, while the BNN includes in addition an uncertainty coming from the limited amount of data provided for training.
Since it is not possible to simulate an arbitrary number of $\gamma$-ray sources according to their true distributions, we have to estimate the uncertainty from a small data set.
The comparison with DNN trained on independent data sets clearly shows that we need to employ BNN to obtain reliable uncertainty estimates.

\subsection{Data augmentation}
A major challenge for the training of machine learning methods is the reduced size of the 4FGL-DR2 data set and the degree of imbalance between the BLL and FSRQ classes.
This imbalance can induce a small bias towards the dominant class in the classification, and has to be taken into account in the evaluation procedure.
One way to handle both imbalanced and small data sets is to use data augmentation, as commonly done in image classification.
In order to evaluate the impact of data augmentation on the estimated uncertainties of a BNN, we employ the popular augmentation algorithm SMOTE~\cite{chawla2002smote} within our toy model.
In its regular setup, additional instances are added between two known sources.
To do so, the distances between all instances of a given class are calculated.
After randomly selecting an instance $S_1$, one of its $k$ nearest neighbors $S_2$ is picked.
A new instance with the same label is then generated at position $S_{new} = S_1 + r (S_2 - S_1)$, with $r \in (0, 1)$ randomly drawn from a uniform distribution.

We test the behavior of the BNN with our toy model, again using the hyperparameters given in table~\ref{tab:param}.
We employ three different setups for training: 
First, we use the BNN with 1000 training points per class as in figure~\ref{fig:correlation_bnn_dnn} as baseline. Second, we train a BNN on 3000 training points per class, out of which 2000 points are sampled according to the SMOTE algorithm.
Finally, we train a BNN on 3000 training points per class that are all sampled from the actual data distributions.

\begin{figure*}[h]
	\includegraphics[trim = 50 50 50 50, clip,width=\textwidth]{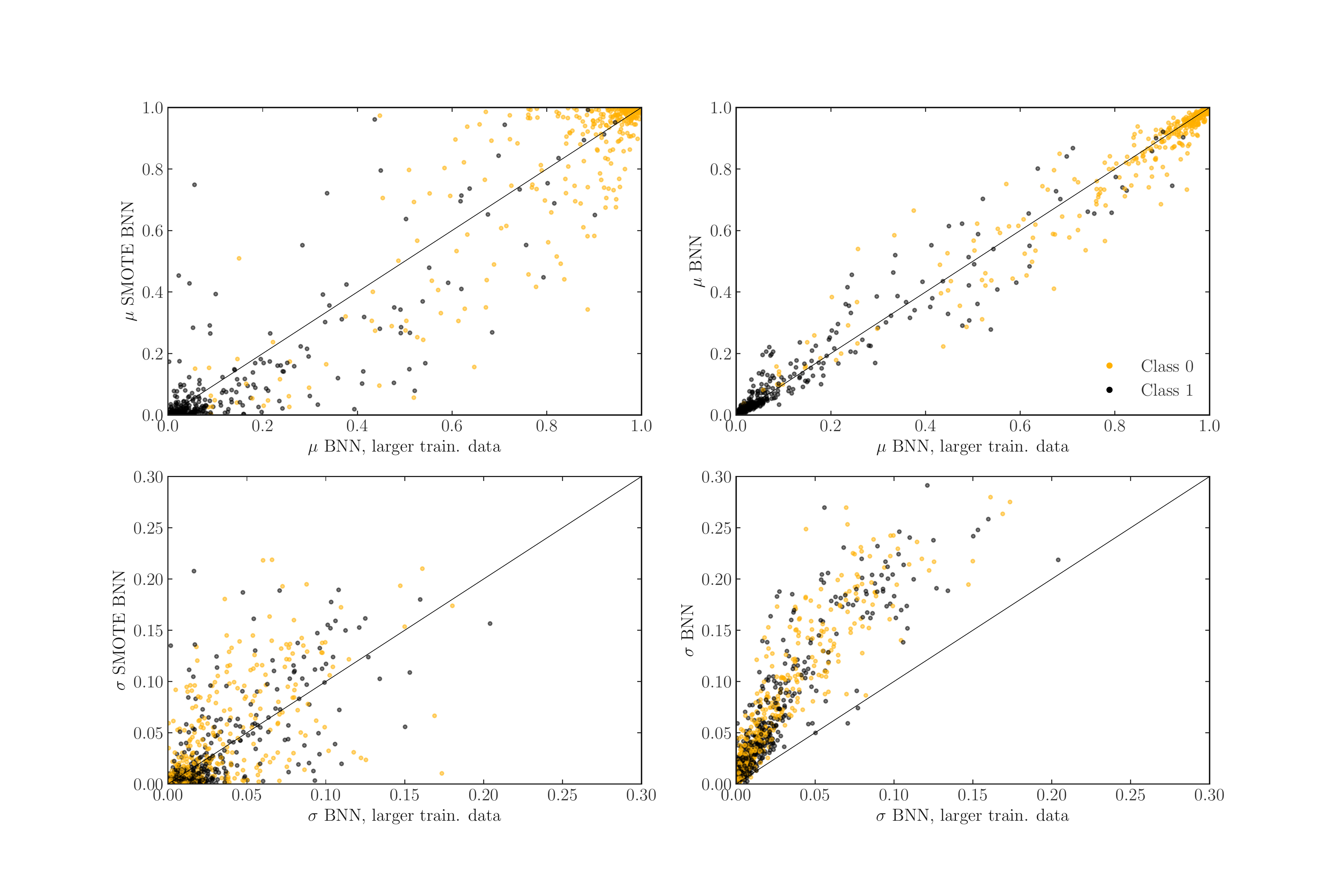}
	\caption{Correlation of the BNN mean predictions ($\mu$, first row) and standard deviations ($\sigma$, second row), exploring data augmentation in our toy model setup. The results of the BNN trained with more samples from the actual distribution (larger train.\ data) is compared to the BNN trained on augmented data using SMOTE (left column) and to the BNN baseline without data augmentation as presented in figure~\ref{fig:correlation_bnn_dnn} (right column).}
	\label{fig:correlation_bnn_augmentation}
\end{figure*}

We show the correlations of predictions and uncertainties for the increased training sample size by additional sampling (larger train.\ data BNN) as compared to augmentation (SMOTE BNN) and to the BNN baseline in figure~\ref{fig:correlation_bnn_augmentation}.
The mean prediction of the SMOTE BNN deviates more strongly from the larger data set BNN than the baseline BNN.
Also, many SMOTE BNN predictions are much closer to zero and one, \ie high probabilities are given for the predicted classes.
As expected, using more data shows a consistent improvement of the uncertainties (lower right panel). 
Instead, using augmented sources (lower left panel) strongly underestimates some uncertainties and overestimates others.

These results show that one needs to be particularly careful with data augmentation when trying to estimate uncertainties.
To find the correct uncertainties within the BNN when trained using data augmentation, one would need to include the assumptions built into the data augmentation procedure into the prior.
This is possible for augmentation by transformations in functional form of individual instances, such as (image) rotations, see \eg refs.~\cite{DBLP:journals/corr/abs-2007-06823}.
However, for algorithms such as SMOTE it is not clear how to do so for our data set.

\section{Results}\label{sec:results}
Following our toy studies we refrain from using data augmentation and employ the BNN to analyze the sources as provided by the 4FGL-DR2 catalog. The goal is to classify different types of blazars and to obtain reliable uncertainty estimates for the network predictions.
We first discuss the classification performance of the BNN on the labeled BLL and FSRQ sources in the 4FGL-DR2 catalog.
We then analyze the impact of different features of the sources on the reliability of the network prediction.
As in ref.~\cite{Finke:2020nrx}, we test the performance of our classifier on a cross-match set of sources between the older 3FGL and the latest 4FGL-DR2 catalog.
In the context of the BNN, this also allows us to assess the accuracy for a given selection of candidates, by using the mean BNN prediction and its standard deviation.
Finally, we focus on the BCU in the 4FGL-DR2 catalog and present two selections for BLL and FSRQ candidates.
The main properties of these candidates are investigated, also in the context of source population studies.

\subsection{Performance}
We first evaluate the classification of the BLL and FSRQ sources in the 4FGL-DR2 catalog. The relatively small sample size requires a careful analysis of the training to avoid effects like overtraining or the dependence on the prior on the latent weight distribution. 
We train our standard BNN architecture with hyperparameters as specified in table~\ref{tab:param} using 10-fold cross validation. We then evaluate the prediction of each source by sampling 500 times from the latent space of the weight distribution. For each source the resulting distribution over predictions can be represented via its mean value and standard deviation. 

\begin{figure}
	\center
	\includegraphics[width=.49\textwidth]{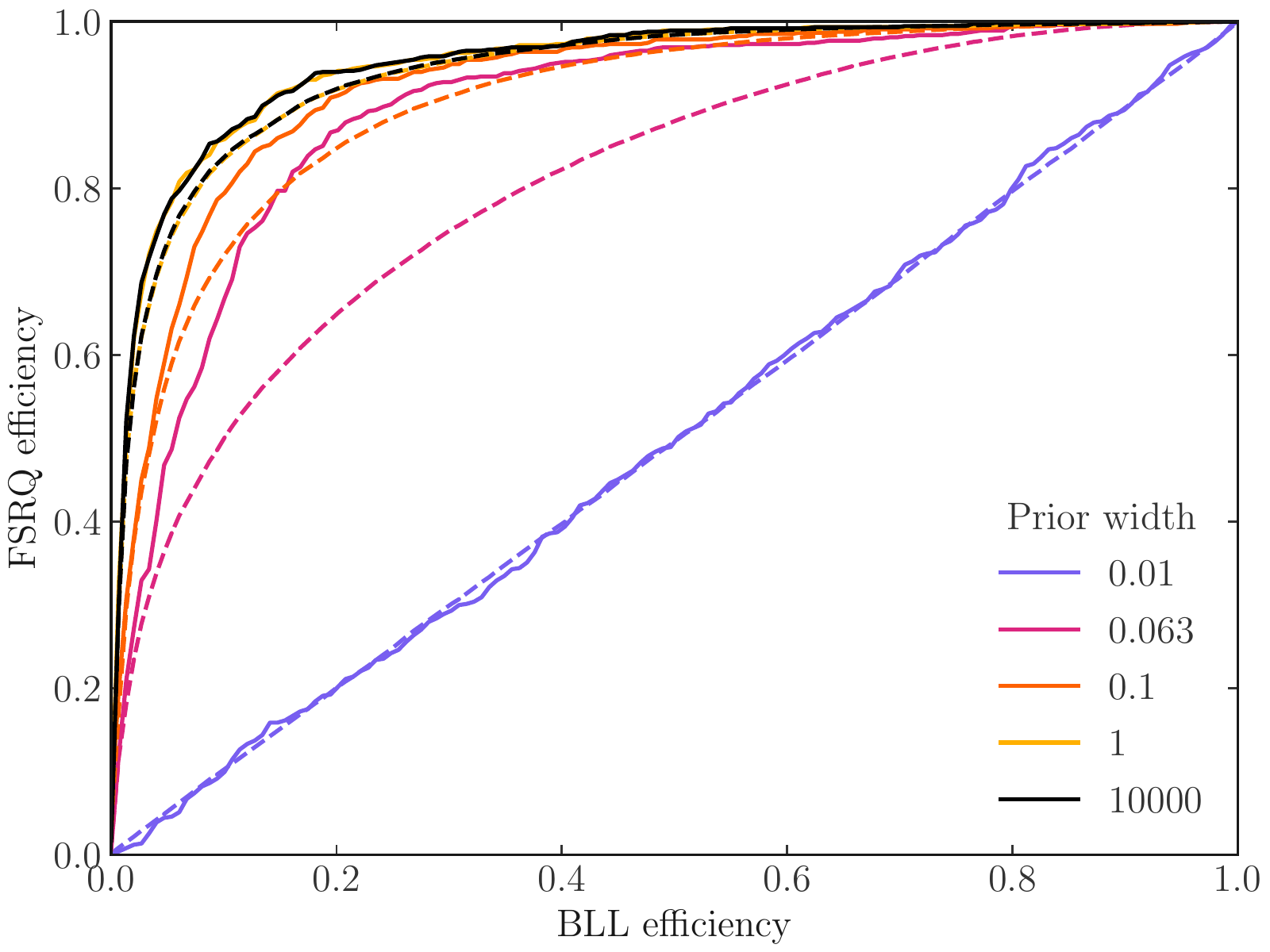}
	\includegraphics[width=.49\textwidth]{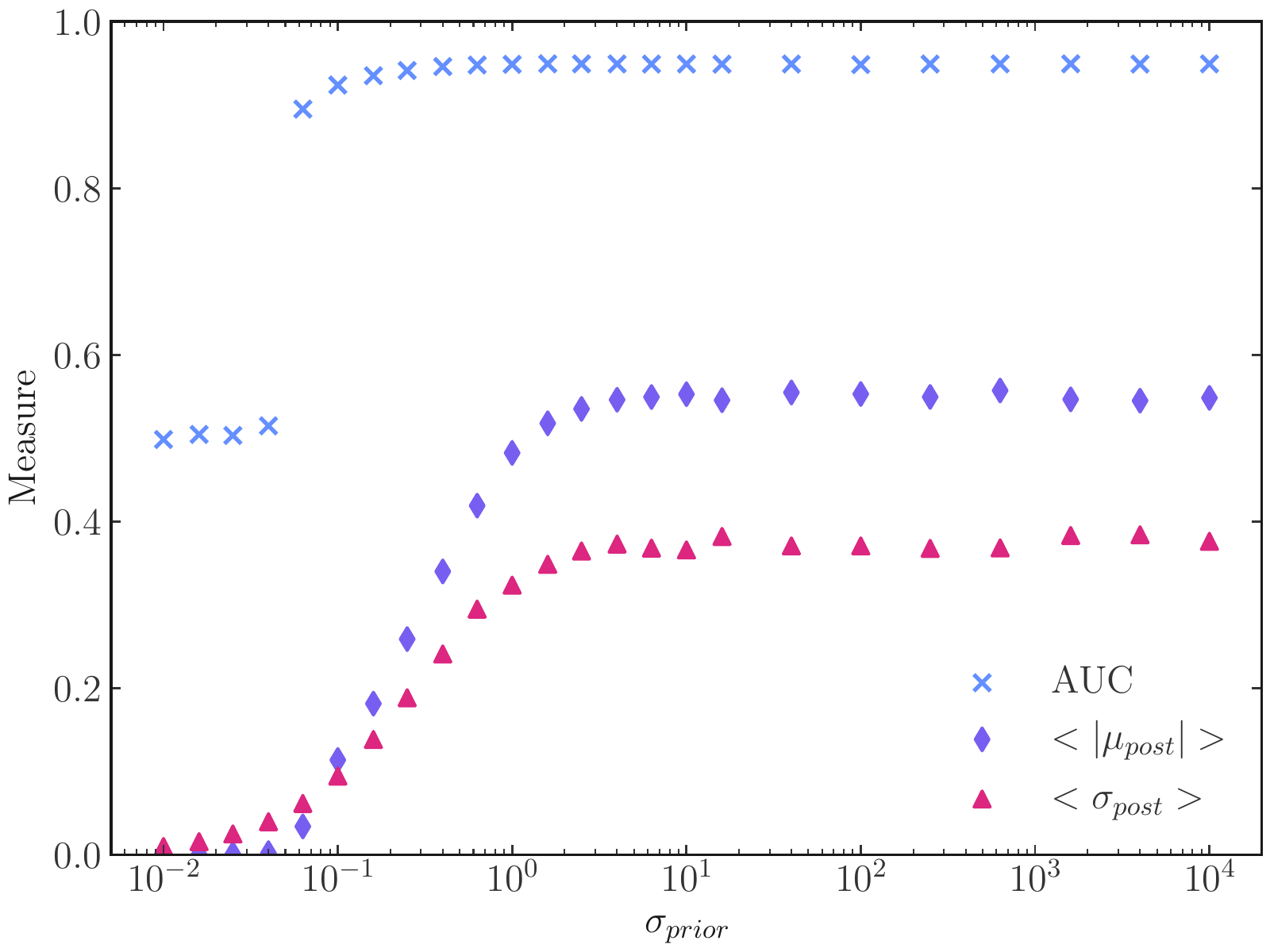}
		\caption{Left panel: ROC curves for the BNN applied to the BLL vs.\ FSRQ classification of 4FGL-DR2 sources using different prior widths.
		The solid lines correspond to the ROC curve of the mean prediction, while the dashed lines correspond to the mean ROC curve over all per prediction ROC curves.
		Right panel: AUC, together with the sum of the absolute values of $\mu_{\rm posterior}$ and the sum of $\sigma_{\rm posterior}$.
		All three measures are normalized such that they have minimum zero and maximum one.}
		\label{fig:bnn_prior}
\end{figure}

The BNN depends not only on the hyperparameters listed in table~\ref{tab:param} but also on the choice of prior over the networks weights.  In the case of large data sets the contribution of the prior dependence to the loss is suppressed by the number of samples as indicated in eq.~\eqref{eq:KL2}.
The smaller the data set, the more carefully we have to assert the impact of the prior. For our training we have chosen Gaussian prior distributions with vanishing mean values and different values for the width.
In figure~\ref{fig:bnn_prior} we show the ROC curves of the mean prediction, the AUC and the mean values of the posterior parameters for training with different choices of the prior width.
For a width below $10^{-1}$ the prior-dependent term enforces very narrow distributions of the weights centered around zero, which dramatically reduces the expressivity of the network. For larger widths the networks is able to learn the underlying features of the data set, leading to an AUC greater than 0.9. The performance saturates for prior widths larger than about one, with only a small remaining prior dependence. We therefore choose $\sigma= 1$ for the following analyses. 

\begin{figure}
	\center
	\includegraphics[width=0.49\textwidth]{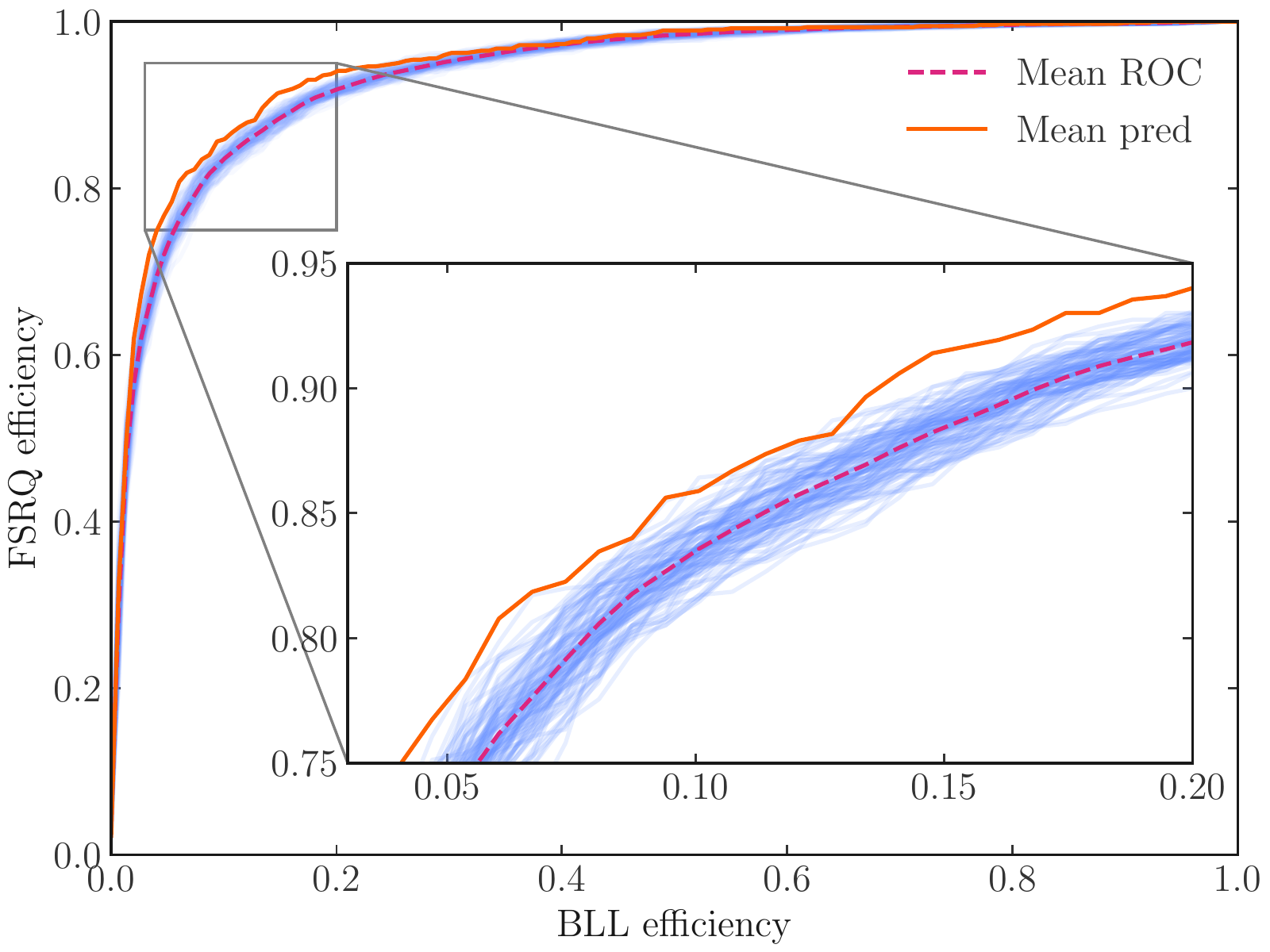}
		\caption{ ROC curves of the BNN applied to BLL vs.\ FSRQ classification of 4FGL-DR2 sources for multiple draws of the weights.
	    The light blue lines correspond to 100 individual draws of the BNN weights, with their mean indicated in magenta.
	    The orange solid line shows the ROC curve after taking the mean of the predictions for each data point.}
		\label{fig:ROC_question}
\end{figure}

Having fixed the prior of the latent distribution, we can analyze the performance of the network. In figure~\ref{fig:ROC_question} we show the final results for the classification of the 4FGL-DR2 sources. The light blue lines illustrate the performance of the BNN for several fixed samples of the network weights. The dashed line indicates the mean of these ROC curves.
The orange line represents the ROC curve once we sample over the latent weight distributions and use the mean prediction for each data point. This sampling procedure averages out the fluctuations that can be observed when we choose fixed weights and therefore leads to an improved performance. This effect can be seen as well when training ensembles of networks.

\begin{figure}
	\includegraphics[width=0.48\textwidth]{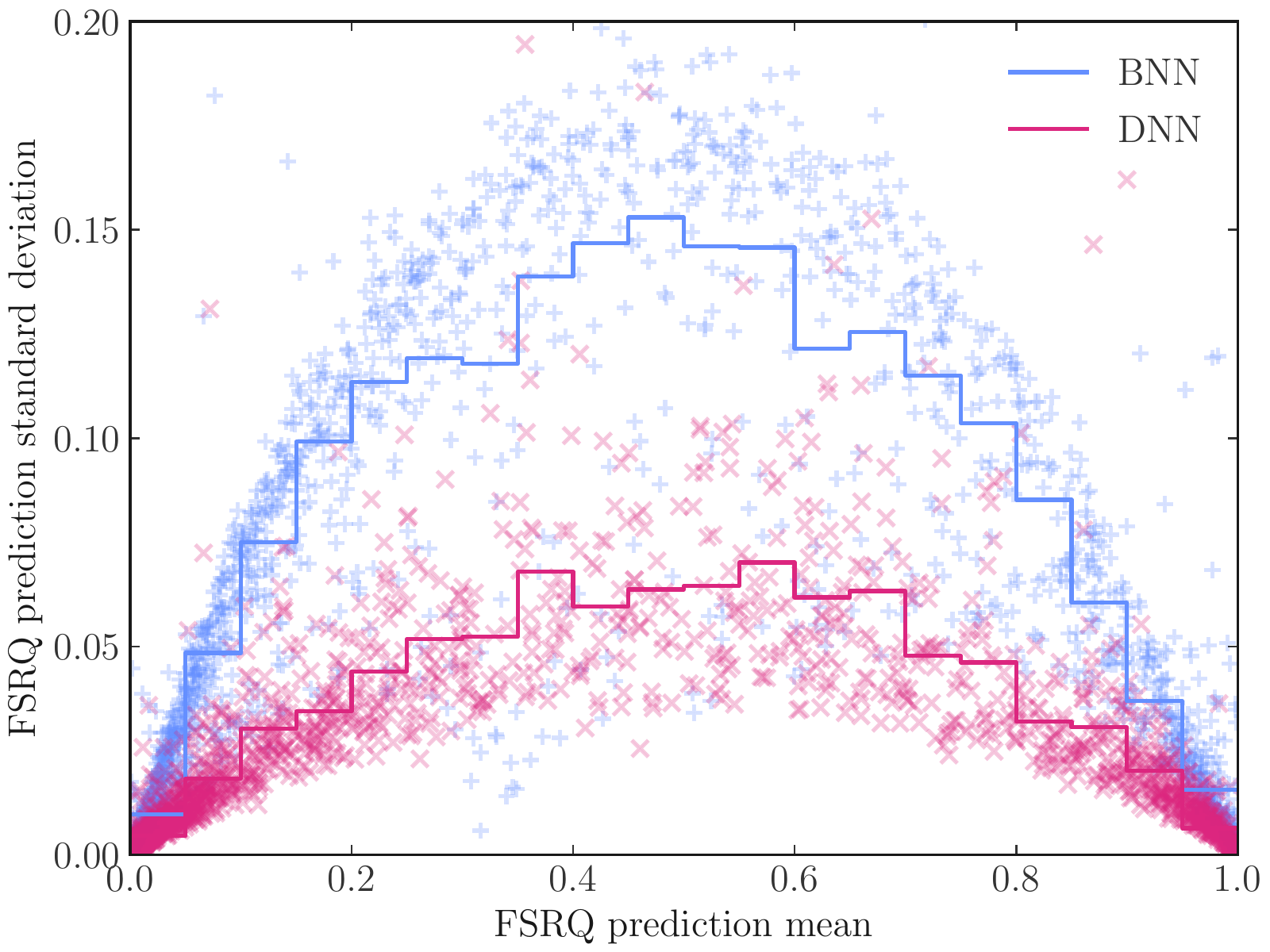}
	\includegraphics[width=0.53\textwidth]{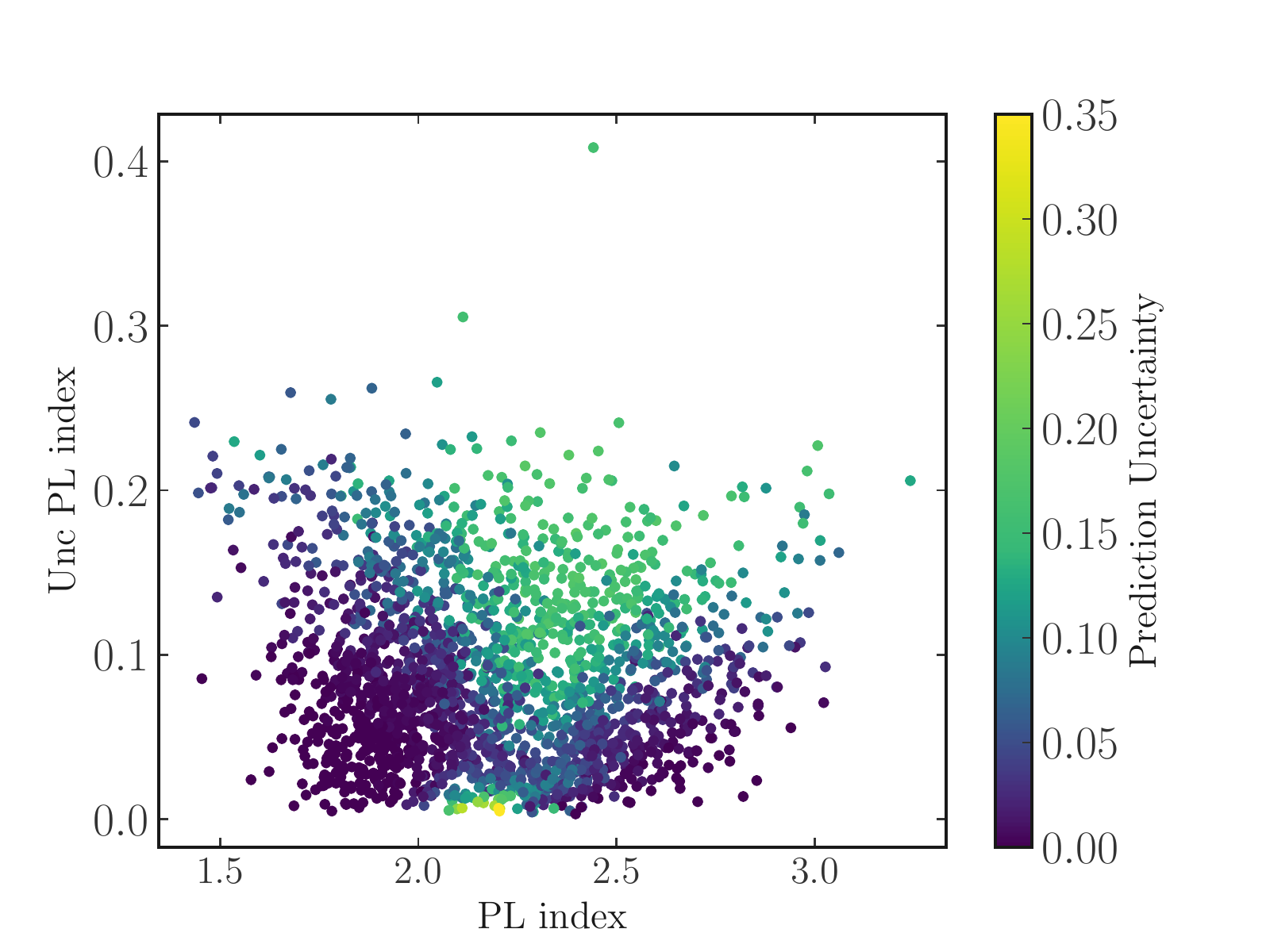}
	\caption{ 
	The left panel shows a comparison of the standard deviation of the BNN (solid blue) and the DNN (dashed magenta) for the mean FSRQ prediction.
	The right panel shows the uncertainty of the BNN predictions for $\gamma$-ray sources of class FSRQ or BLL. They are reported in the plane spanned by the power law index and its associated uncertainty, as given in the 4FGL-DR2, to identify the characteristics of sources with large BNN uncertainties.}
	\label{fig:calibration_rainbow}
\end{figure}

When using BNN for classification we need to distinguish the mean prediction $\mu$ for a source to belong to a certain class from the uncertainty on this prediction $\sigma$. 
In the left panel of figure~\ref{fig:calibration_rainbow} the FSRQ standard deviation is plotted against the mean FSRQ (positive label) prediction for 100 draws of the BNN weights (blue) and for the DNN ensemble (magenta).
The scatter points correspond to individual sources, while the histograms show the mean value within each bin for the BNN and the DNN, respectively.
In particular for spectra with $0.4 <\mu < 0.6$, where the network considers both labels equally likely, we observe that the associated uncertainties of the BNN ranges from 0.04 to 0.2.
Although the mean predictions $\mu$ of the classifiers are calibrated correctly for DNN and BNN (see appendix), the standard deviations of the ensemble of 100 DNN (trained on the same training set, compare with section~\ref{sec:methods}) clearly underestimate the uncertainty of their prediction.
This result corroborates what we observed within our toy model, and strengthens again the case for using BNN for our classification task.

Within the BNN we can explore for which sources the network assigns larger or smaller prediction uncertainties, and if this is correlated to some characteristics of their energy spectrum. %
While we do not use any derived feature as given in \fermi catalogs as input for the DNN and BNN, these features can be a useful tool to understand what the network actually learns from the energy spectrum, as demonstrated for the AGN vs.\ PSR classification in ref.~\cite{Finke:2020nrx}. 
In the right panel of figure~\ref{fig:calibration_rainbow} we plot the data set in the plane spanned by the power law index of the sources of the training set (without distinguishing between FSRQ and BLL) and its associated uncertainty. The color map depicts the BNN prediction uncertainty. 
We observe that the BNN prediction uncertainty is smaller around the central value of the BLL ($\sim2$) and FSRQ ($\sim2.5$) power law distributions. 
The BNN assigns larger uncertainties when predicting the label of sources with properties in between the two distributions (power law index $\sim2.2$), which confirms the observation already outlined for the left panel of figure~\ref{fig:calibration_rainbow}.
Moreover, we find that the BNN assigns larger uncertainties to sources with large uncertainties of the power law index, \ie fluctuating energy spectra. 
Finally, we have verified that the sources with the highest BNN uncertainties (yellow dots) have energy spectra outside the typical flux range. This is expected, since out of distribution samples correspond to regions with low statistics, so that the network is less likely to converge on a prediction.

\subsection{Cross-match with the 3FGL catalog}\label{sec::crossmatch}
Some sources that were not labeled in the previous version of the Fermi-LAT $\gamma$-ray catalog have now been associated or identified in the 4FGL-DR2. 
This allows us to test the performance of the BNN on a cross-match data set using the 3FGL catalog.
Moreover, since this subset contains mainly fainter objects with respect to the bulk of the training set, it allows to test a potential sample selection bias of the supervised classifier, see discussion in refs.~\cite{Finke:2020nrx,Luo_2020}.
We thus train the BNN on the subset of sources which have a common classification in the two catalogs (653 BLL and 458 FSRQ, \ie 1111 sources in total), and use as a test set the newly classified sources in the 4FGL-DR2 (647 BLL and 283 FSRQ, \ie 930 sources in total). Unlike to what was done in ref.~\cite{Finke:2020nrx}, we here use always the energy spectrum as given in the 4FGL-DR2 catalog. 
For training and testing we follow the same architecture and procedure as described in section~\ref{sec:methods}.
Using exclusively the mean prediction of the BNN to classify sources as BLL ($\mu<0.5$) and FSRQ ($\mu>0.5$), we obtain an accuracy of 85.5\%.
This value is comparable to the accuracy of 88.9\% obtained within the 4FGL-DR2 sources, confirming the robustness of the algorithm. 

The cross-match test further allows us to define new selection criteria which can be later applied to classify BCU sources. We define a \emph{tight} and a \emph{loose} selection to either achieve a high accuracy for the prediction, or to obtain labels for a large number of classified sources. 
By using both the $\sigma$ and $\mu$ from the BNN, we define the classification threshold $\tau$ such that a source is classified as FSRQ if the mean predicted class probability minus its standard deviation surpasses a given threshold, \ie $\mu -\sigma< \tau$. 
For the BLL class, this corresponds to $\mu +\sigma< 1- \tau$. 
 
The number of sources in the cross-match data set that are classified for various thresholds, and the corresponding accuracy of the classifications are reported in table~\ref{tab:threshold_selection}. 
We choose as loose threshold a value of 0.5, which maximizes the number of classified sources and which achieves an accuracy estimated to be $93.31\%$. 
The tight threshold is set instead to 0.8, resulting in a higher accuracy of $96.33\%$ but with about half of the sources classified.
The corresponding confusion matrices are illustrated in the appendix~\ref{sec:appendix}. 

While we can estimate the accuracy of each selection threshold within this cross-match setup, this is not possible when applying the BNN to the full set of BCU in the 4FGL-DR2 catalog. However, we expect that the loose and tight selections defined here will produce sets of candidates with comparable accuracy to what is shown in table~\ref{tab:threshold_selection}.

\begin{table}[]
    \centering
    \caption{Number of sources  that are classified for various thresholds $\tau$ in our cross-match test with the 3FGL, together with the estimated accuracy. We choose as \emph{loose} threshold the one maximizing the number of classified sources, while the \emph{tight} threshold is the one maximizing the obtained accuracy, \ie 0.5 and 0.8 respectively.}
    \begin{tabular}{ccc}
    \toprule
        Threshold ($\tau$)  & $N_{Sources}$ & Acc [\%]  \\
         \midrule
        {\bf 0.5}   & 703           & 93.31     \\
         0.6        & 577           & 93.93     \\
         0.7        & 455           & 95.16     \\
         {\bf 0.8}  & 327           & 96.33     \\
         0.9        & 194           & 95.88     \\
        \bottomrule
    \end{tabular}
    \label{tab:threshold_selection}
\end{table}

\subsection{Predictions for BCU in the 4FGL-DR2 catalog}\label{sec:pred}
\begin{figure}
	\includegraphics[width=0.5\textwidth]{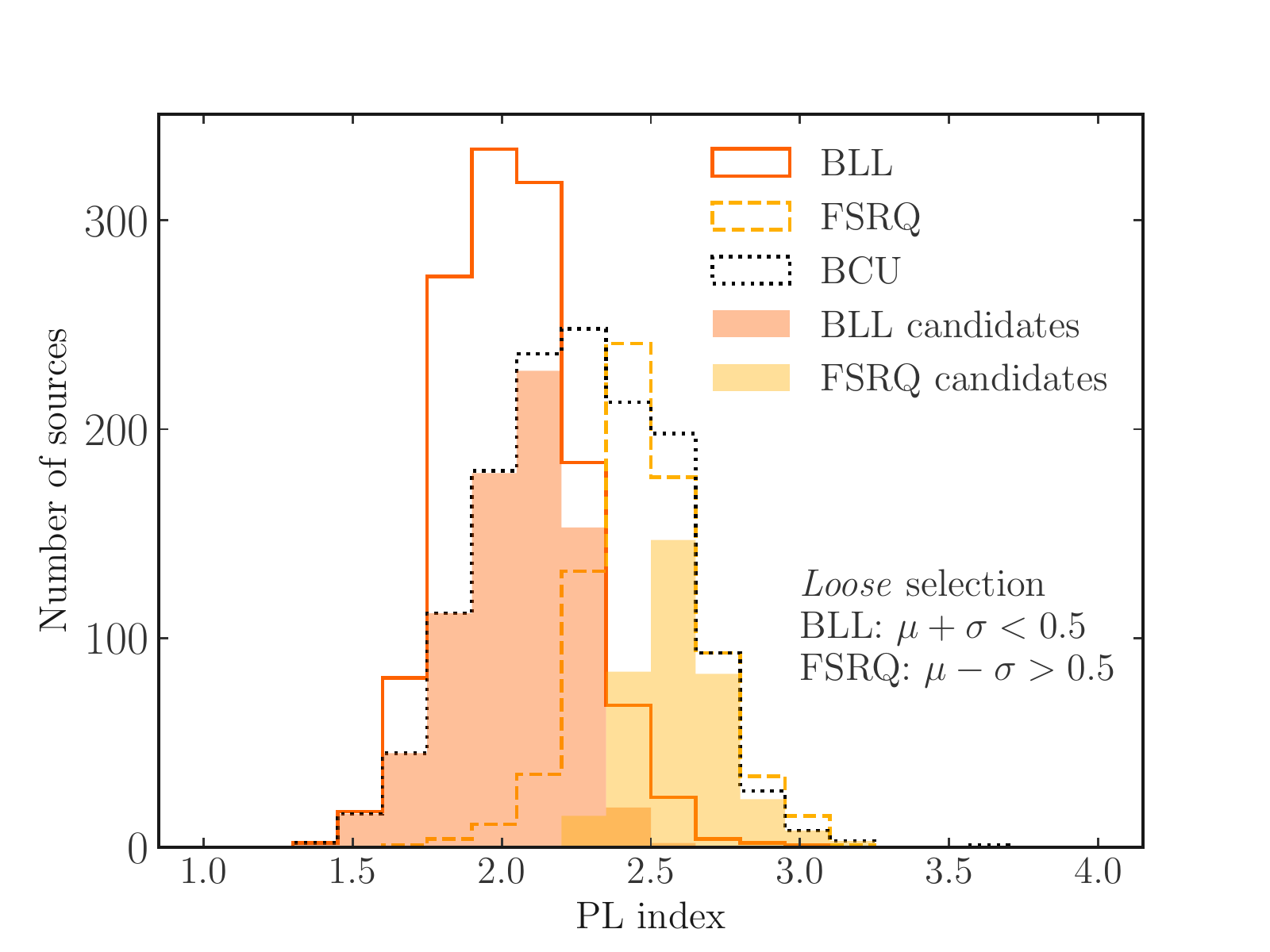}
	\includegraphics[width=0.5\textwidth]{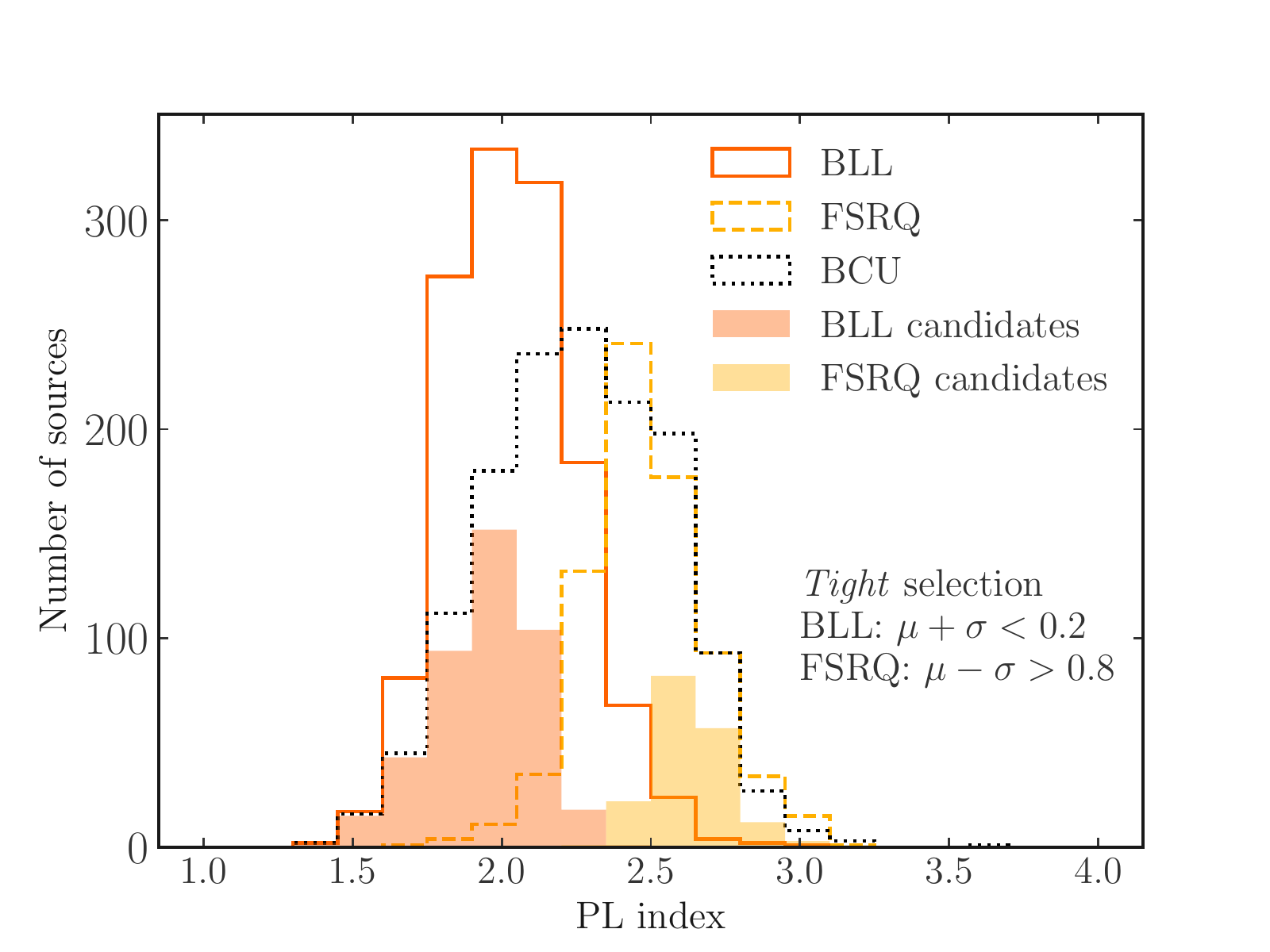}
		\caption{The hollow histograms depict the power law index distribution for BLL, FSRQ and BCU in the 4FGL-DR2 catalog. The filled histograms depict the BLL and FSRQ candidates obtained with the BNN applied to the BCU. 
		In the left (right) panel we illustrate the candidates as selected using the loose (tight) selections, in which both the $\mu$ and the $\sigma$ of the BNN prediction are used. The threshold values defining the loose and tight selections are also reported in each panel for reference.}
		\label{fig:pred_hist}
\end{figure}

Using the loose and tight selection criteria derived in section~\ref{sec::crossmatch} we now apply the BNN to the BCU in the 4FGL-DR2 catalog. 
For the loose selection we obtain 756 BLL and 363 FSRQ candidates. In the left panel of figure~\ref{fig:pred_hist} we show the distribution of the power-law index for the candidate sources. The BLL and FSRQ candidates are centered around a value of about 2.0 and 2.5, respectively, in excellent agreement with the distributions of the labeled sources, which are shown as hollow histograms in the same panel.
For the tight selection we obtain a smaller subset of classified sources with 429 BLL and 178 FSRQ candidates. Comparing the two selections in figure~\ref{fig:pred_hist} we see that the tighter selection mainly removes candidate sources in the intermediate region of the power-law index, while leaving the tails of the distribution mostly unchanged. This is in agreement with the distributions shown in the right panel of figure~\ref{fig:calibration_rainbow}, where we found that sources with large uncertainties are concentrated in regions with intermediate values of the power-law index.
The resulting distributions are still peaked roughly around the same central values, moving just slightly further apart. 
Overall, the FSRQ and BLL candidates extracted from the BCU have the expected spectral characteristics as derived from the labeled sources.

Apart for producing a list of promising, individual targets for follow-up observations, the results of our classifications can be used to complement population studies aimed at estimating the collective properties of $\gamma$-ray blazars. 
One of the main application of population models is to quantify the contribution of different astrophysical source classes to the extragalactic $\gamma$-ray background emission \cite{Massaro:2015gla,Zechlin:2016pme,Roth:2021lvk}, and thus constrain exotic contributions, such as from dark matter (see ref.~\cite{2015PhR...598....1F} for a review). 
Moreover, constraining the properties of blazar populations has also important consequences for the study of IceCube neutrinos, since AGN are among the main candidates to produce high energy astrophysical events, see \eg refs.~\cite{Palladino:2018lov,Abbasi:2021jyg}. 

Due to the increasing number of blazars detected by \textit{Fermi}-LAT, their properties can be studied by tuning population models using $\gamma$-rays only, without relying on extrapolations based on lower frequencies. 
To improve statistics, FSRQ and BLL are often considered together in unified population models \cite{Ajello:2015mfa,Manconi:2019ynl}. 
However, comparing models in which the two classes are considered jointly or separately shows that the model predictions for the number of sources are not well constrained, in particular at low $\gamma$-ray fluxes \cite{Ajello_2012,Ajello_2013}.

To further investigate the population of FSRQ and BLL candidates, we compute the source-count distribution $dN/dS$, \ie the number of sources $N$ as a function of the $\gamma$-ray source flux $S$. 
At large photon fluxes, the source-count distribution is expected to have large statistical fluctuations, as the number of sources is relatively small. At low fluxes, the apparent decrease of the source-count distribution is caused by a decrease in the detection efficiency \cite{DiMauro:2017ing}. 

\begin{figure}
\center
	\includegraphics[width=0.49\textwidth]{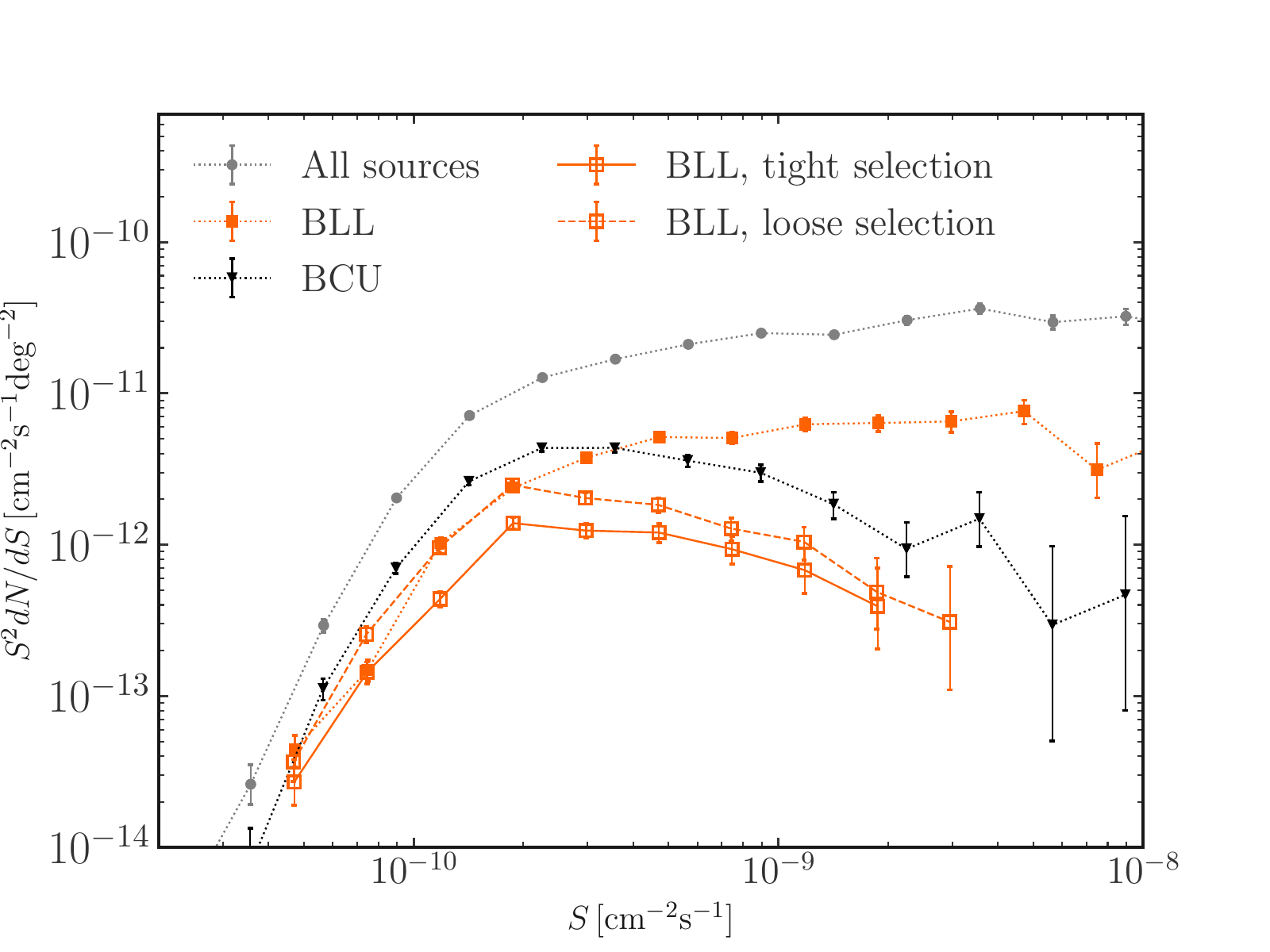}
	\includegraphics[width=0.49\textwidth]{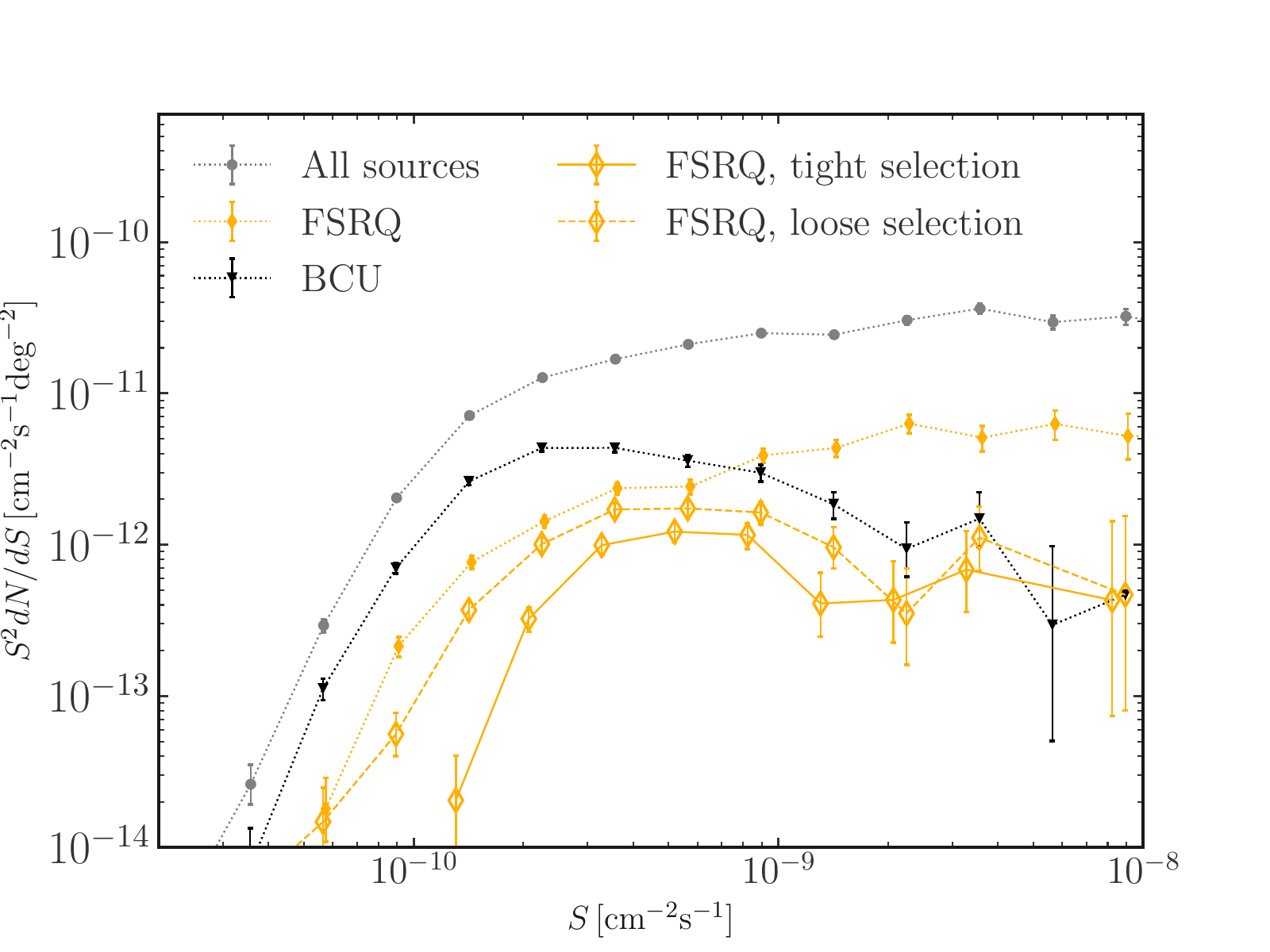}
		\caption{ Source count distribution (computed using $\gamma$-ray fluxes between 1--100 GeV) for all the sources in the 4FGL-DR2 catalog, together with individual contributions from BLL, FSRQ and BCU (filled gray, orange, yellow and black markers, respectively). 
		The loose and tight BNN predictions for the BLL (FSRQ) are reported in the left (right) panel with empty markers. 
		For each point, errors depict the Poisson error; points are connected with lines to guide the eye.
		}
		\label{fig:pred_dNdS}
\end{figure}

In figure~\ref{fig:pred_dNdS} we show the source-count distribution for BLL (left panel) and for FSRQ (right panel) blazars using the $\gamma$-ray flux computed in the energy range from $1$ to $100$~GeV.  For each set of sources, the values are computed assuming five bins per $\gamma$-ray flux decade, with Poisson error. Lines connecting the $dN/dS$ values are superimposed to guide the eye. 
The $dN/dS$ of all the sources in the catalog 
(gray) and of the BCU (black) is also added in each panel for comparison.
We first note that blazars indeed constitute a significant part of the observed sources across all the $\gamma$-ray flux interval, as also evident from the numbers in figure~\ref{fig:classtree}. 
The BCU distribution is peaked at low fluxes near the detection threshold of the catalog, where new sources have been recently discovered and have not yet been identified or associated using multi-wavelength observations.
In each panel, the BLL and FSRQ candidates obtained within the tight and loose selections are shown, and compared to the distribution of the labeled sources. 
Focusing on the loose selection, we note that the BNN identifies more BLL candidates among the BCU in the low flux regime ($S\sim 10^{-10}$ cm$^{-2}$ s$^{-1}$ deg$^{-2}$), while almost all of the BCU at high fluxes ($S\sim 10^{-9}-10^{-8}$ cm$^{-2}$ s$^{-1}$ deg$^{-2}$) are classified as FSRQ candidates. 
This nicely confirms the trend already visible within the labeled sources in the catalog, and follows the expectations from the current models of the FSRQ and BLL luminosity evolution, which predict the BLL to be overall more numerous at low $\gamma$-ray source fluxes \cite{Ajello_2012,Ajello_2013,Lisanti:2016jub}.
Similar conclusions can be drawn from the candidate sources within the tight selection, although they are overall less numerous, in particular at lower fluxes. 
By summing the labeled sources in the catalog and the sources selected by the loose cuts, we predict that $60\%$ ($32\%$)  of $\gamma$-ray blazars in the 4FGL-DR2 catalog are BLL (FSRQ). 
The cumulative source count distribution obtained by the sum of labeled and candidates BLL, FSRQ sources could be compared to the extrapolation of blazar models fitted to the labeled sources only.  Also, our tight candidate sources could be included in future model fits to complement the statistics of sources in the low-flux regime. 
We leave these analyses to future investigations. 

We provide the complete list of FSRQ and BLL for the two selections in the github repository:  \url{https://github.com/manconi/agn-psr-nn-classification}.

\section{Summary and conclusions}\label{sec:conclusions}
We have devised a Bayesian neural network (BNN) architecture for the classification of $\gamma$-ray sources as detected by \emph{Fermi}-LAT. We have focused in particular on blazars, which constitute the majority of identified sources and which can be classified further into BL Lacertae objects (BLL) and Flat Spectrum Radio Quasars (FSRQ). 
Our neural networks have been trained on the energy spectrum of identified blazar sources as provided by the 4FGL-DR2 \fermi catalog. As the data set available for training is comparatively small, standard dense neural networks (DNNs) are susceptible to overtraining and may generalize poorly. We have thus employed BNNs, which are more robust and which provide a more reliable estimate of the uncertainties on their prediction.

To quantify the difference in performance between standard and Bayesian neural networks, and to explore the effect of data augmentation, we have conducted a study based on a toy data set sampled from a multivariate Gaussian distribution. We compare the classification probabilities of the Bayesian neural network with those of two ensembles of standard networks that have been trained repeatedly with the same toy data set and with toy training data sets sampled independently, respectively. We observe a strong correlation of the mean prediction and the standard deviation between the BNN and the ensemble of DNNs trained on independent training data. Moreover, training an ensemble of DNN on the same data set significantly underestimates the uncertainties. 

The small and imbalanced data set provided by the \fermi catalog is a generic challenge for machine learning methods, which might be alleviated by data augmentation. We employ the commonly used SMOTE algorithm to quantify the effect of data augmentation within our toy model setting. We find that training the BNN using augmented sources leads to unreliable estimates of classification uncertainties. Within the Bayesian setting, the assumptions built into the data augmentation procedure should be included in the prior on the weight distributions. While this may be possible for data augmentation that relies, for example, on the symmetries of the data set, it is not clear how to modify the prior in general for our specific data set. 

For the classification of  blazars of uncertain type  in the most recent \fermi catalog, we have trained a BNN on the energy spectrum of identified BLL and FSRQ sources. For the reasons mentioned above, we have not performed any data augmentation. We have verified that the performance of the BNN is largely independent of the specific choice of hyperparameters and the choice of prior. We have provided various performance measures such as ROC curves, accuracies, calibration curves as well as the confusion matrix. To verify the results from our toy model study we have compared the predictions of the BNN with those obtained from an ensemble of DDNs trained repeatedly on the same data set. As expected, the standard deviation as derived from the ensemble of DNNs significantly underestimates the uncertainty of the classification prediction. 

We find that the BNN assigns larger uncertainties to those blazar candidates that have a spectral index in between the average values found for BLL ($\Gamma_{\gamma} \sim 2$) and FSRQ ($\Gamma_{\gamma} \sim 2.5$), or that have strong fluctuations in the energy spectra. The sources with the largest uncertainties possess spectra with fluxes outside the typical range, corresponding to regions in the training data set with particularly low statistics. 

We can further test the performance of the BNN through a classification of those $\gamma$-ray sources that were  unclassified in the previous version of the \fermi catalog, but have now been classified as blazars through multiwavelength observations. Training the BNN on the subset of sources which have a common classification in the two catalogs, and using the newly classified sources in the 4FGL-DR2 catalog as test set, we obtain a classification accuracy of about 85\%, which is comparable with the accuracy obtained by using sources within the recent catalog only.

For the final classification of the  blazar candidate objects in the 4FGL-DR2 catalog we define a tight and a loose selection. A source is classified as FSRQ if the mean predicted class probability minus its standard deviation surpasses 0.8 (tight selection) or 0.5 (loose selection), respectively. The loose selection maximizes the number of classified sources and achieves an accuracy of about 93\%, while the tight selection results in a higher accuracy of about 96\%, but with approximately half of the sources classified. We find that the tighter selection essentially removes candidate sources that have a spectral index in between the average values found for BLL and FSRQ, respectively. 

A reliable classification of thus far unclassified blazar candidates is important input to population studies, which aim for example to quantify the contribution of different astrophysical source classes to the extragalactic $\gamma$-ray background emission. We have compared the  source count distribution, \ie the number of sources as a function of the $\gamma$-ray flux, of blazars identified through multiwavelength observations and classified by our tight and loose selection. We find that the BLL candidates classified by the BNN are more abundant at low $\gamma$-ray fluxes, consistent with the behavior of the identified sources and the expectations from current models of the BLL and FSRQ luminosity evolution. 

In summary, we have shown that Bayesian neural networks provide a robust classifier of blazar candidate $\gamma$-ray sources with a reliable uncertainty estimate. Bayesian neural networks are particularly well suited for classification problems that are based on small and imbalanced data sets. Such data sets are common, not only in $\gamma$-ray astrophysics, and we expect Bayesian neural networks to become a standard for classification tasks in astroparticle physics in general.

\section*{Acknowledgements}
We are grateful to Kathrin Nippel for discussions and comments on the manuscript. TF is supported by the Deutsche Forschungsgemeinschaft (DFG, German Research Foundation) under grant 400140256 - GRK 2497: The physics of the heaviest particles at the Large Hadron Collider. 
The research of AB and MK is supported by the Deutsche Forschungsgemeinschaft (DFG, German Research Foundation) under grant 396021762 - TRR 257: Particle Physics Phenomenology after the Higgs Discovery.

\bibliography{astroml_ep2}
\bibliographystyle{JHEP}

\appendix

\section{Additional results}\label{sec:appendix}
\subsection{Calibration}
\begin{figure}
	\center
	\includegraphics[width=0.5\textwidth]{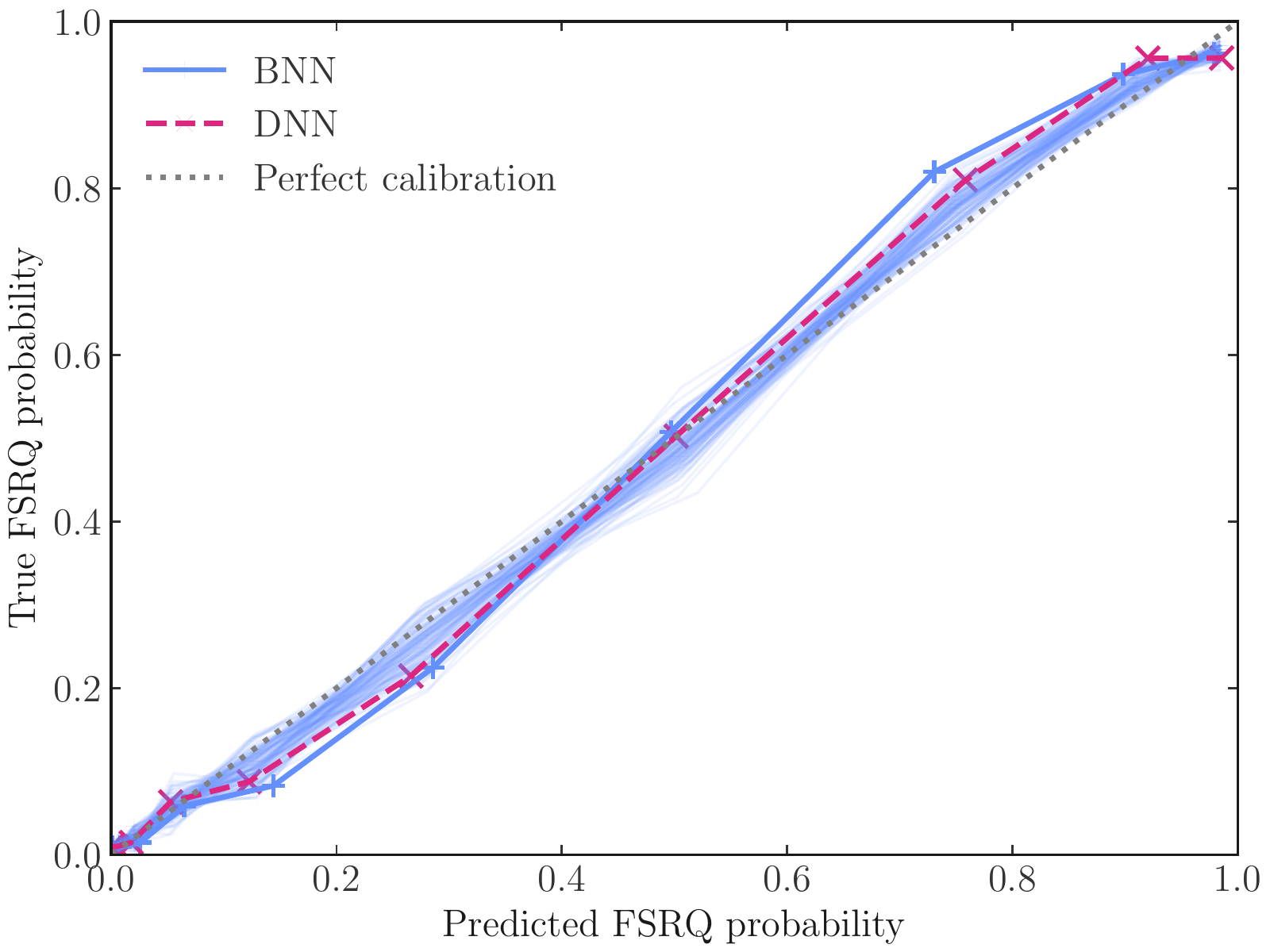}
	\caption{Calibration curve for the BNN (blue solid) and the DNN (magenta dashed).
	A sample of 100 draws from the BNN is also shown as light blue solid lines.
	The bins along the x-axis are chosen such that each bin contains the same number of sources.}
	\label{fig:calibration}
\end{figure}

The calibration of our classifiers is illustrated in figure~\ref{fig:calibration}, by comparing the predicted FSRQ probability to the true FSRQ probability for the BNN (blue crosses, connected by a blue line for better visibility) and the DNN (magenta crosses and dashed line).
For the BNN both the mean calibration curve (thick line) as well as a sample of 100 draws (thin lines) are shown.
To calculate the points for each classifier, the predictions are separated into ten equally populated bins. For each bin we determine the $x$-value by calculating the mean prediction (FSRQ probability). The corresponding $y$-value is then given by the fraction of FSRQ sources within that bin. 
A perfectly calibrated classifier should produce points along the diagonal (dotted line).
We observe that both the BNN and the DNN are well calibrated.

\subsection{Confusion matrix}
To complement the performance measures of our BNN classifier, we present in table~\ref{tab:conf_mat_xCat} the confusion matrix for the cross-match test performed in section~\ref{sec::crossmatch}, and in  table~\ref{tab:conf_mat_1lay} for the classification of BLL vs.\ FSRQ in the 4FGL-DR2 catalog. We show results obtained with the tight and loose selections described in the main text.  
The confusion matrix for binary classification is a $2\times2$ matrix.
The entry $ij$, with $i$ denoting the row and $j$ the column, shows the number of sources with true label $i$ and predicted label $j$.
In our specific case, the number of true BLL or FSRQ which are correctly predicted by the network then corresponds  to the diagonal entries. 
The off-diagonal entries show instead the number of sources which are misclassified by the network, \ie the number of BLL which are wrongly classified as FSRQ and vice versa.
A perfect classifier would give non-zero entries only on the diagonal.
As expected, the tight selection  minimizes the number of misclassifications to less than about $4\%$ of the total sources; for the loose selection the corresponding figure is $8\%$.

\begin{table}[h!]
    \caption{Confusion matrix for classifying the cross catalog sources using a threshold of 0.8 (left) or 0.5 (right).}
    \label{tab:conf_mat_xCat}
    \centering
    \begin{tabular}{lcc}
        \toprule
                    & Pred BLL  & Pred FSRQ \\
         \midrule
         True BLL   & 261       & 7         \\
         True FSRQ  & 5         & 54        \\
        \toprule
    \end{tabular}
    \hspace{0.8cm}
    \begin{tabular}{lcc}
        \toprule
                    & Pred BLL  & Pred FSRQ \\
        \midrule
         True BLL   & 518       & 15        \\
         True FSRQ  & 32        & 138       \\
        \toprule
    \end{tabular}
\end{table}

\begin{table}[h!]
    \caption{Confusion matrix for classifying the 4FGL-DR2 sources using 10 fold cross validation and a threshold of 0.8 (left) or 0.5 (right).}
    \label{tab:conf_mat_1lay}
    \centering
    \begin{tabular}{lcc}
        \toprule
                    & Pred BLL  & Pred FSRQ \\
         \midrule
         True BLL   & 861       & 16        \\
         True FSRQ  & 25        & 342       \\
        \toprule
    \end{tabular}
    \hspace{0.8cm}
    \begin{tabular}{lcc}
        \toprule
                    & Pred BLL  & Pred FSRQ \\
        \midrule
         True BLL   & 1142      & 53        \\
         True FSRQ  & 81        & 544       \\
        \toprule
    \end{tabular}
\end{table}

\end{document}